\documentclass{llncs}
\usepackage{cite}
\usepackage[pdftex]{graphicx}
\usepackage[caption=false,font=footnotesize]{subfig}
\usepackage{url}
\usepackage{comment}

\usepackage{listings}
\usepackage{mathrsfs}
\usepackage{times}
\usepackage{color}
\usepackage{ulem}
\newcommand{\implem}[1]{\textbf{\texttt{#1}}}
\newcommand{\rdc}[0]{radio duty-cycling}
\begin{document}

\title{Technical Report : ContikiMAC performance analysis}

\author{Mathieu Michel \and Bruno Quoitin}

\institute{Computer Science Department,
University of Mons, Belgium\\\texttt{firstname.lastname}@umons.ac.be}
\maketitle

\begin{abstract}
ContikiMAC is a radio protocol that allows nodes to communicate while
keeping their radio turned off most of the time. The radio duty-cycle
of ContikiMAC can be as low as 1\%. To signal an incoming frame to a
receiver, ContikiMAC repeatedly sends the full data frame until it is
acknowledged  by the receiver. This approach might seem less efficient
than what older \emph{radio duty cycle} protocols have been using. X-MAC for example
first sends short \emph{strobe} frames to signal the receiver of an incoming 
frame. It is only when those strobes are acknowledged that the actual
data frame is transmitted.

In this paper, we perform an in-depth study of the ContikiMAC
protocol to understand why sending full data packets works well in
practice. We use the X-MAC protocol as a baseline. In addition to the
genuine X-MAC protocol, we also experiment with enhanced versions of
X-MAC that include some of the ContikiMAC mechanisms. This allows to
better quantify which mechanism is responsible for what performance
enhancement.
Both protocols performances are evaluated in terms of expected transmission count (ETX), latency, packet delivery ratio (PDR) and
duty-cycle. Our study reveals that the better efficiency of ContikiMAC
can be attributed to two specific mechanisms: first the ``fast sleep"
optimization that shortens the wake-up period and second a more efficient
transmission procedure. The combination of these mechanisms helps
ContikiMAC to achieve a better PDR than X-MAC together with a reduced latency and a
drastically lower energy consumption.
A preliminary version of those results was published in French\cite{macomp}. 
This paper extends those results, in particular by performing comparison on larger topologies and using RPL, as routing protocol, instead of static routing
\end{abstract}

\section{Introduction}

During the last decade, several solutions have been proposed to
address WSN requirements and constraints such as limited bandwidth
and low energy consumption. In particular, a lot of research has been
dedicated to avoiding situations where nodes unnecessarily keep their
radio active while there is no incoming packet to
receive~\cite{sampling-survey}. Such situation, known as
\emph{idle-listening}, has been solved by the design of \emph{radio duty cycle}
(RDC) protocols at the link layer. These protocols force every node to
periodically switch its radio transceiver between short active
(\textit{listen}) periods and long inactive (\textit{sleep})
periods. The duty cycle is the ratio between the duration of the
active period and the interval between two successive wake-ups; the shortest the duty cycle, 
the lowest the energy consumption.

There are two main categories of \emph{radio duty cycle} protocols :
\emph{asynchronous} and \emph{synchronous} protocols. Synchronous
duty-cycle protocols force the nodes to synchronize their wake-up
periods while asynchronous protocols let every node free to determine
its own independent sleep schedule. By removing the need to create,
share and maintain a common active period, asynchronous
protocols are less complex to implement and lead to lower protocol
overhead. Asynchronous protocols are therefore well suited to low bandwidth
networks where they are particularly energy
efficient~\cite{sampling-survey}. Several asynchronous protocols have
been described in the literature such as LPL~\cite{lpl},
B-MAC~\cite{bmac}, WiseMAC~\cite{wisemac}, X-MAC~\cite{xmac} and
more recently ContikiMAC~\cite{contikimac}. X-MAC and ContikiMAC are
among the most popular due to their availability in the Contiki RTOS~\cite{contiki-paper:2004}. We focus on these protocols in
this paper.

ContikiMAC and X-MAC differ in their transmission
procedures. To signal an incoming frame to the receiver, X-MAC first
sends a stream of short-sized strobes. When the receiver acknowledges
one of those strobes, the data frame is sent. ContikiMAC uses a
different approach : it repeatedly transmits the full data packet until it
is acknowledged by the receiver. Sending full data packets might a priori
appear as less energy efficient, but surprisingly ContikiMAC has been
reported to outperform X-MAC \cite{contikimac}. ContikiMAC allows
duty-cycle as low as 1\% and has a wake-up mechanism 10 times more
efficient than that of X-MAC. Other studies show that ContikiMAC is
able to drastically reduce the energy consumption and to obtain a
better latency at the expense of strict timing constraints and higher noise 
sensitivity~\cite{ccathreshold,contikimac,perf}.

To better understand the efficiency of ContikiMAC we perform
controlled experiments in the Contiki simulator
COOJA~\cite{cooja:2006}. The objective of our study is to quantify
the gain that can be obtained by sending full data packets instead of
starting with strobes. We use X-MAC as a baseline. To this end, we
first identify the main mechanisms involved in ContikiMAC and then
design new versions of X-MAC where some of those mechanisms are
incorporated. The main changes brought to X-MAC are in the wake-up and
frame transmission procedures. We then compare the performance of
ContikiMAC against X-MAC and its derivatives on several scenarios.

This paper, is organized as follows. We first introduce the X-MAC and
ContikiMAC protocols in Section~\ref{sec:background1}. Then, we
describe in Section~\ref{sec:methodology} our test methodology, the enhancements brought to X-MAC and
the experimental setup. We present
and explain the results of our experimentations without routing in
Section~\ref{sec:results-regular-topo} and with RPL routing protocol in 
Section~\ref{sec:results-random-topo}. Finally we conclude and discuss
possible future works in Section~\ref{sec:conclusion1}.

% ===================================================================
\section{Background}
\label{sec:background1}

X-MAC and ContikiMAC are two asynchronous \rdc{} protocols. No common
wake-up schedule is established a priori among the nodes but instead,
each time a frame must be transmitted, the sender needs to synchronize
with the receiver. The approaches used by X-MAC and ContikiMAC are
described in the following subsections. Although they are intimately
linked, the packet transmission mechanisms are described in
Section~\ref{sec:bg-pkt-tx} and the node wake-up mechanisms are
described in Section~\ref{sec:bg-wake-up}.

% ------------------------------
\subsection{Packet transmission}
\label{sec:bg-pkt-tx}
The role of the packet transmission mechanism consists for a
transmitting node in determining when a neighbor destination node is
ready to receive a data frame. We explain below the details of the
approaches used by X-MAC and ContikiMAC and highlight their
differences.

\subsubsection{X-MAC}

\begin{figure}[ht!]
  \begin{center}
    \includegraphics[scale=0.4]{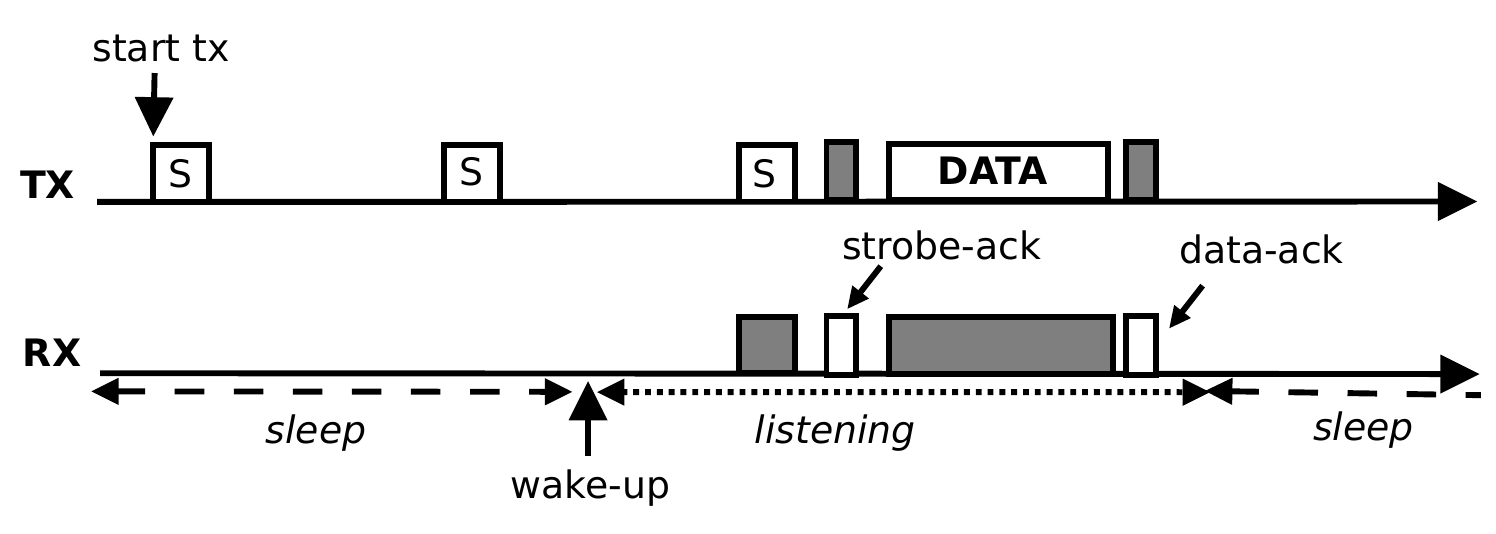}
    \caption{X-MAC uses a stream of strobes to advertise a transmission.}
    \label{fig:xmac}
  \end{center}
\end{figure}

X-MAC uses different mechanisms to unicast or broadcast a frame.
The unicast transmission mechanism derives from the preamble
sampling technique introduced by B-MAC~\cite{bmac} and LPL~\cite{lpl}.
In this approach, receivers are notified of an incoming frame by
the transmission of a long preamble, the duration of which is greater
than the wake-up interval. X-MAC replaces the use of such a long preamble
by a stream of short \emph{strobe} frames. When the intended receiver catches 
one strobe it replies with a \emph{strobe-ACK}. The sender then stops sending strobes and 
proceeds with the data frame transmission. This mechanism is depicted in
Fig.~\ref{fig:xmac} for a successful transmission.
%The first two strobe frames (S) are sent while the receiver (Rx) is
%asleep. The third strobe is acknowledged and followed by the data frame.

A transmission will fail if any of the following conditions is met :
(a) No strobe-ACK is received after a duration equivalent to a
wake-up interval, or (b) No data-ACK is received after the data frame
transmission. In either cases, it is the responsibility of the above
layers to schedule a retransmission.

Compared to B-MAC and LPL, this methodology reduces overhearing and
latency, resulting in lower energy consumption. 
\begin{enumerate}
  \item \textbf{Overhearing reduction} is achieved by embedding
    the destination address in the strobes. Receivers can check
    this address to determine if they need to stay awake or if they
    can return to sleep immediately.
    
  \item \textbf{Latency reduction} is achieved by inserting a short delay between consecutive strobes to
    allow the destination neighbour to send a \emph{strobe-ACK}. When a strobe-ACK is
    received, the stream of strobes is interrupted and the data packet
    is sent immediately. 
\end{enumerate}
 
The broadcast transmission mechanism of X-MAC also relies on a stream of
strobes. The difference with unicast is that strobes contain no
destination address : every nearby receiver is concerned. Moreover, to
allow all the receivers to wake-up, the strobe stream has a duration
slightly longer than a wake-up interval. No strobe-ACK is sent by receivers to
acknowledge the reception of strobes.

% - - - - - - - - - - - - - - - - - - - - 
\subsubsection{ContikiMAC}

\begin{figure}[ht!]
  \begin{center}
    \includegraphics[scale=0.4]{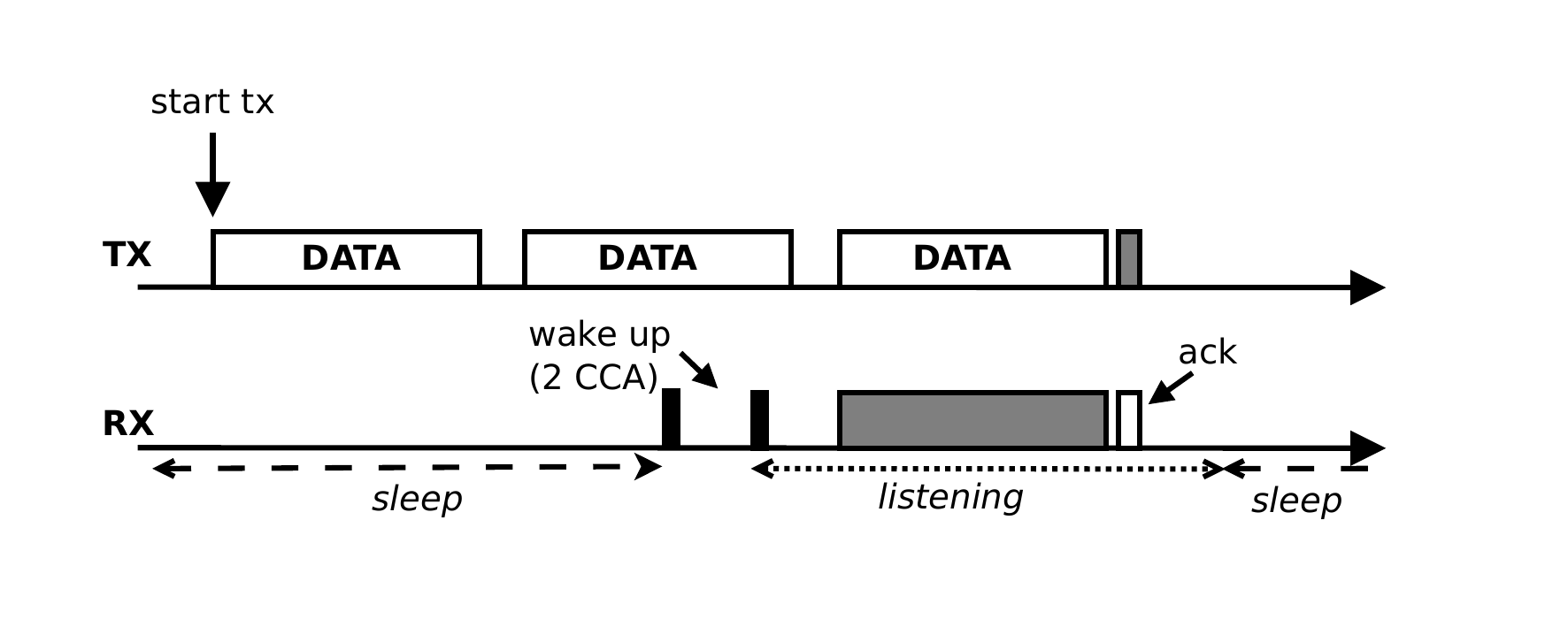}
    \caption{ContikiMAC repeatedly sends the data packet until an ACK is received.}
    \label{fig:ContikiMAC}
  \end{center}
\end{figure}

To send a packet, a node running ContikiMAC repeatedly sends the full
data frame. The frame destination field allows to reduce overhearing :
a node that is not the destination of the frame can immediately go
back to sleep. In the opposite case the receiver acknowledges the
correct reception of the frame. When an ACK is received, the sender
stops sending the data frame and the transmission is successful. This
mechanism is illustrated in Fig.~\ref{fig:ContikiMAC}.

A transmission will fail if no ACK is received after a duration
  equal to a wake-up interval. In this case, it is the responsibility
  of the above layers to schedule a retransmission.

Broadcast transmissions are achieved in the same way than unicast, except that no data-ACKs are expected.

% ---------------------------------------
\subsection{Wake-up procedure}
\label{sec:bg-wake-up}

The role of the wake-up mechanism is to determine when a node needs to
wake-up and when it can go to sleep. The simplest approach is to
wake-up periodically for a fixed duration and then sleep for another
fixed duration. However X-MAC and ContikiMAC can decide to go to sleep earlier and hence save
additional energy. They do this in conjunction with their transmission mechanisms. 
We detail in the following subsections the approaches followed by X-MAC
and ContikiMAC and highlight their differences.

\subsubsection{X-MAC}

X-MAC forces each node to wake up at regular interval for a short \emph{active} period. During its active period, a node listens for potential incoming packets. If no incoming transmissions are detected the node goes back to sleep at the end of its active period and stays asleep until the next scheduled wake-up.
The length of an active period is typically 5 or 10\% of the wake-up interval. A typical wake-up interval value is 125~ms. The active period can be extended beyond the cycle length if the node is still involved in transmissions at the end of the cycle.

% - - - - - - - - - - - - - - - - - - - - 
\subsubsection{ContikiMAC}

A node running ContikiMAC also needs to wake up periodically. However, to the contrary of X-MAC, a node does not need to stay awake until the end of its active period. The node quickly performs two successive Clear Channel Assessments (CCA) to determine, based on the radio signal strength, if there is an incoming transmission. If the CCAs succeed (channel clear), the node goes back to sleep immediately. If the CCAs fail (channel busy), the node stays awake and a procedure called a \emph{fast sleep optimization} tries to determine if the received signal is due to noise or to an incoming frame. In case of noise the node goes back to sleep.

\begin{figure}[ht!]
  \begin{center}
    \includegraphics[scale=0.4]{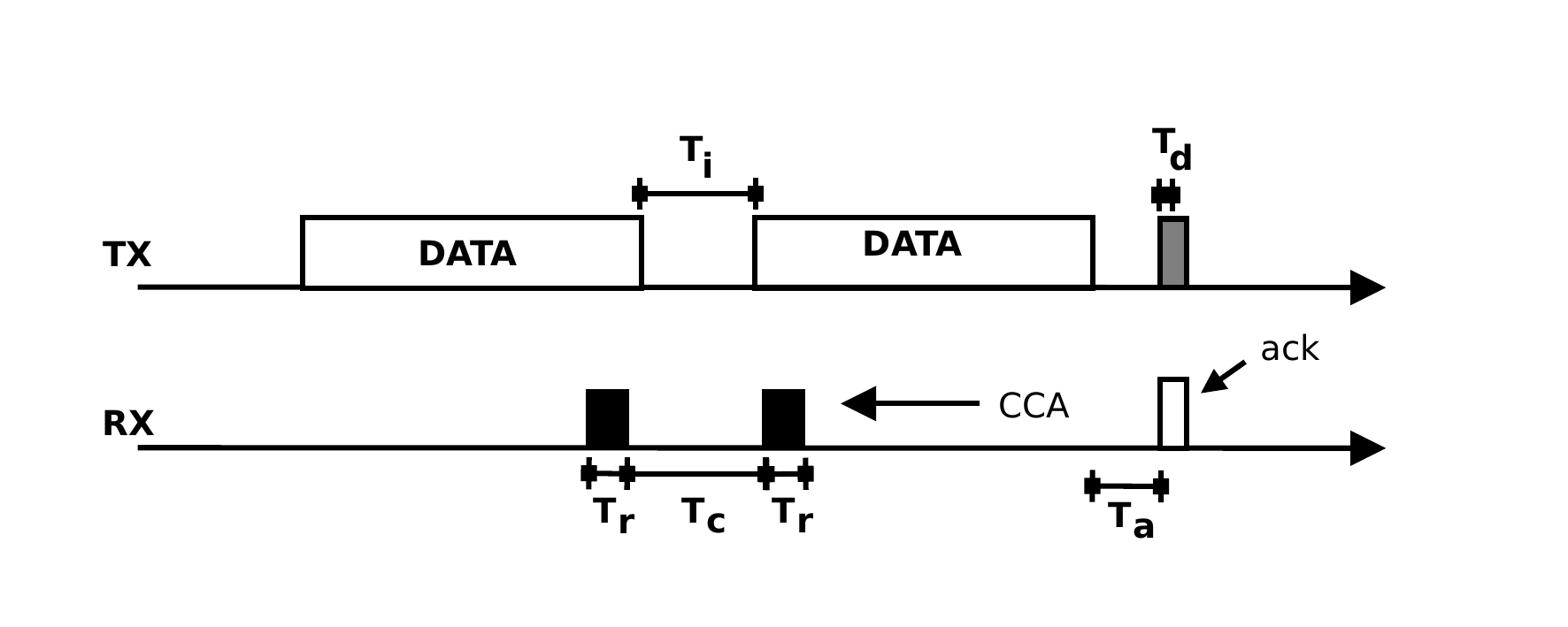}
    \caption{Timing constraints in ContikiMAC (inspired from ~\cite{contikimac})}
    \label{fig:ContikiMAC-timing}
  \end{center}
\end{figure}

The timing of the wake-up phase is critical to make data transmission functional. Compared to an X-MAC node which stays awake for a specific time, the use of CCAs by ContikiMAC implies very strict timing constraints\cite{contikimac}. A summary of those constraints are defined below. The variables used in the constraints are described in Table~\ref{tab:cmac-timing-notations} and an illustration is provided in Fig.~\ref{fig:ContikiMAC-timing}.

\begin{enumerate}

  \item \textbf{Space for ACKs} : The interval between two transmissions ($T_i$) must be greater than the time needed to send and receive an ACK ($T_a + T_d$).
  
  \item \textbf{Detect frame with only 2 CCAs} : To ensure that at worst two CCAs are enough to detect a transmission, the interval $T_c$ between two CCAs must be longer than the interval between two transmissions $T_i$.
  
  \item \textbf{Minimum frame duration} : To avoid that a full transmission occurs between two CCAs, the shortest packet duration ($T_s$) must not be smaller than $T_r + T_c + T_r$ 
  
\end{enumerate}

\begin{table}[t!]
\begin{center}
\begin{tabular}{|c|p{9cm}|}
\hline
\textbf{Variable} & \textbf{Description} \\
\hline
$T_i$ & Interval between two consecutive data frames. \\
$T_r$ & Time required to perform a single CCA. \\
$T_c$ & Interval between two consecutive CCAs. \\
$T_a$ & Time delay between the end of a data frame and the start of the ACK frame.\\
$T_d$ & Time required to detect an ACK. \\
$T_s$ & Transmission time of the shortest packet.\\
\hline
\end{tabular}
\caption{Variables used in ContikiMAC timing constraints.}
\label{tab:cmac-timing-notations}
\end{center}
\end{table}

% \begin{comment}
% The following notations are used :
% \begin{itemize}
%   \item $T_{i}$ interval between two data packet transmissions.
%   \item $T_{r}$ time required to perform a single CCA.
%   \item $T_{c}$ interval between two CCAs.
%   \item $T_{a}$ time delay between the end of de data frame and the start of the ACK transmission.
%   \item $T_{d}$ time required to detect the ACK.
%   \item $T_{s}$ transmission time of the shortest packet.
% \end{itemize}
% \end{comment}

Those constraints may be summarized by $T_{a}+T_{d} < T_{i} < T_{c} < T_{c}+2T_{r} < T_{s}$

% One might wonder if it would make sense to implement the \emph{fast sleep optimization} in X-MAC ? The answer is No. given that the time interval between two strobes ($T_{sInt}$ = 3.9 ms) is by default 4 times larger than the interval between two ContikiMAC successive data transmissions ($T_{i}$ = 0.9ms). A substantial increase of the number of successive CCAs needed to sense the medium would be required to cover an interval of time long enough to detect a transmission.

% ---------------------------------------
\subsection{Additional ContikiMAC features}

The performance of ContikiMAC is improved by two additional mechanisms.

\begin{figure}[ht!]
  \begin{center}
    \includegraphics[scale=0.4]{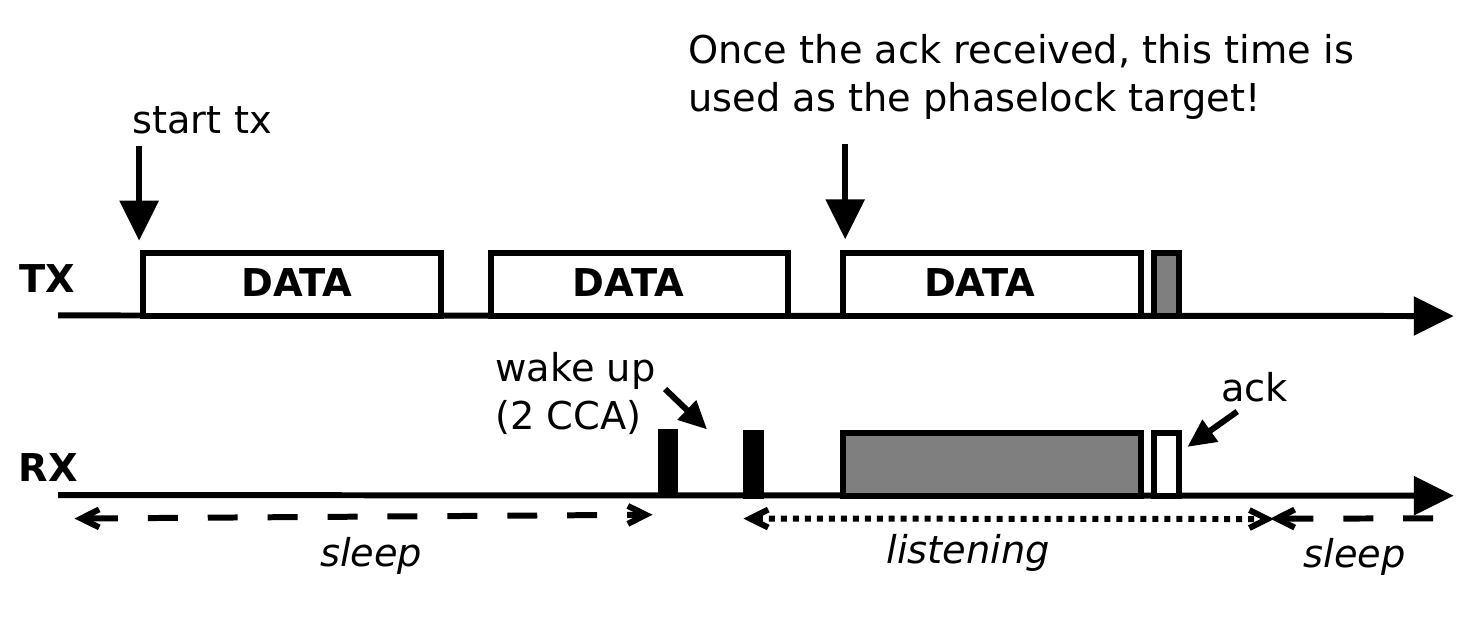}
    \caption{When an ACK is received, the sender estimates the
      wake-up schedule of the receiver.}
    \label{fig:ContikiMACP1}
  \end{center}
\end{figure}

\begin{figure}[ht!]
  \begin{center}
    \includegraphics[scale=0.4]{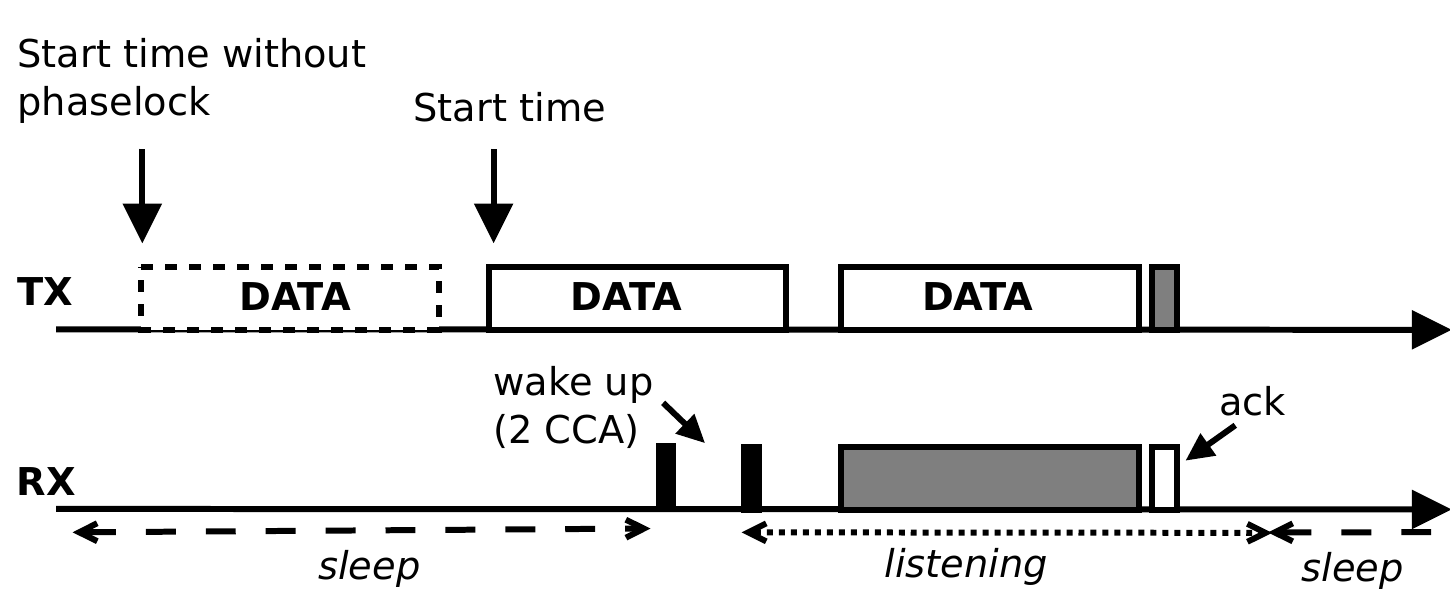}
    \caption{When the neighbour schedule is known, the length of the transmission is reduced.}
    \label{fig:ContikiMACP2}
  \end{center}
\end{figure}

First, a \textbf{phase-lock mechanism} allows a node to learn the wake-up phase of a neighbour. When a node receives an ACK it learns that the destination is awake at this moment (Fig.~\ref{fig:ContikiMACP1}).
 The sender can then make use of this knowledge during the next transmission towards the same neighbour. It can start its repeated data frame transmissions just before the receiver will \textbf{supposedly} wake up (Fig.~\ref{fig:ContikiMACP2}). This mechanism allows a node to achieve a transmission to a ``phase-locked neighbor '' with only two packets on average (the first being used to announce the transmission). Therefore the phaselock mechanism decreases the time and the energy spent in TX mode. A side benefit of the phase-lock is a reduction in channel utilization and therefore in the risk of collisions. Such a mechanism was first introduced in WiseMAC~\cite{wisemac}

Second, a \textbf{collision avoidance mechanism} checks the
availability of the channel before data transmission by performing several successive CCAs. This is done in a way such that the interval
covered by the CCAs is slightly longer than the time $T_{i}$ between two successive data
packet transmissions . The lack of activity during this period
informs a node that apparently no other transmission is
occurring. If one CCA fails (channel busy), the sender will postpone its transmission. This mechanism prevents a node to start a transmission
while another one is ongoing, avoiding a collision.

% -----------------------------------------------------------
\subsection{X-MAC implementation in Contiki}
\label{sub:sec: contikix-mac}

The Contiki implementation of X-MAC \cite{contiki-xmac} that we use
in this study slightly differs from the original version
\cite{xmac}. This section briefly describes the main differences with
the original version.

\begin{enumerate}

  % \item \textbf{Broadcast:}  The broadcast mechanism uses an alternative approach, similar to ContikiMAC, which consists of repeated transmissions of the data packet. This broadcast mechanism is proposed by default in Contiki. The number of repetitions must be defined in such a way that each potential receiver has the possibility to wake up to catch the packet. This number must be large enough to cover a full wake-up interval.
  
  \item \textbf{Encounter optimization:} The implementation of X-MAC
    provided by Contiki contains a mechanism similar to the phase-lock
    of ContikiMAC. This mechanism should allow a sender to learn the
    wake-up agenda of a receiver and, as a consequence, make a
    transmission possible with only two strobes on average. However,
    we found out that in practice the implementation of this mechanism
    blocks other possible transmissions while waiting for the receiver
    wake-up. For this reason, we disabled it. We discuss a retrofit of the ContikiMAc's phaselock mechanism in X-MAC in Section~\ref{sec:methodology}.

%    \textbf{This optimization, which to the best of our knowledge has not been validated, should have similar results than the ContikiMAC phaselock mechanism. In practice the results of our tests are disappointing due to a non-optimal design/implementation.}
   % In practice the results are disappoiting due to a non-optimal implementation of this encounter optimization involves a blocking wait before meeting an encounter: if a node wants to send a data packet to a node for which the wake-up schedule is established, the sending process is paused until the destination's wake-up. This prevents the node to handle other tasks.}
  
  \item \textbf{Reliable data transmission:} Contiki's X-MAC implementation uses a flag which is set to indicate the need for a data packet acknowledgement. Given the use of the strobe-ACK, the original X-MAC paper~\cite{xmac} considers data ACK as optional. Contiki's X-MAC allows to request a data-ACK for more robust transmissions. For such transmissions the sender stays awake waiting for the ACK. This ACK can be sent without requiring any strobes since the transmitter is already awake.
  
  \item \textbf{Streaming:} When enabled, a node doesn't need to send a strobe before each packet transmission in a same data flow. This mechanism is similar to the ``burst mode''in ContikiMAC allowing a node to send several packets in a row.

  \item \textbf{Collision avoidance:} Contiki's X-MAC implementation does not provide an efficient collision avoidance
    mechanism : to the opposite of ContikiMAC, no Clear Channel Assessment(CCA) is performed before a
    transmission. However, X-MAC will postpone a pending packet transmission if any strobe
transmission has been detected during the current active period.
  
\end{enumerate}

% ===================================================================
\section{Methodology}\label{sec:methodology}

%The objective of this paper is to understand why ContikiMAC performs better than X-MAC while it uses a mechanism which seems a priori less bandwidth efficient. As shown in Section~\ref{sec:background1}, the main differences between ContikiMAC and X-MAC are in the frame transmission and node wake-up procedures.

This section describes the methodology used to compare ContikiMAC and X-MAC. We aim to better understand the mechanisms or the combination of mechanisms used in ContikiMAC's transmission and wake-up procedures. To reach this objective we use the X-MAC protocol as a baseline. In addition to the genuine X-MAC protocol, we also consider enhanced versions of X-MAC that include a CCA based collisions avoidance mechanism and a phaselock mechanism similar to ContikiMAC. We then compare the performance of these different versions in the same scenarii.

% -----------------------------------------------------------
\subsection{Protocols under evaluation}

Table~\ref{tab:mac} summarizes the 5 protocols that we use in our evaluation. \implem{ContikiMAC} and \implem{X-MAC} are the genuine implementations of those protocols in Contiki. Variations of X-MAC are also considered where CCA collision avoidance (\implem{C}) and/or phase-lock (\implem{P}) are added separately or together. This leads to three additional protocols : \implem{X-MAC-C} where only the CCA collision avoidance has been added; \implem{X-MAC-P} where only the phase-lock mechanism has been added; and \implem{X-MAC-CP} where both mechanisms were added.

\begin{table}[ht!]
\begin{center}
    \begin{tabular}{|l|c|c|}
      \cline{2-3}
      \multicolumn{1}{c|}{} & \textbf{CCA Col. Av.} & \textbf{Phaselock} \\
      \hline
      \implem{ContikiMAC} & X & X\\
      \hline
      \implem{X-MAC} &  & \\
      \hline
      \implem{X-MAC-C} & X & \\
      \hline
      \implem{X-MAC-P} &  & X\\
      \hline
      \implem{X-MAC-CP} & X & X\\
      \hline
    \end{tabular}
    \vspace{0.5cm}
    \caption{Summary of the protocols under evaluation}
    \label{tab:mac}
    \end{center}
\end{table}

% ===================================================================
\subsection{Simulation setup}
\label{sec:simulation-setup}

We perform simulations with the Contiki 2.6 COOJA
Simulator~\cite{cooja:2006}. COOJA is a Java-based discrete-event
simulator. It allows to perform simulations at the network and
firmware levels. The main advantage is that simulations can be run
with almost the same code as deployed on real platforms. Several
protocol implementations such as ContikiMAC, X-MAC and RPL are
available in the Contiki source code. The experiments are based on the
Zolertia Z1 platform~\cite{zolertia} which is well supported by
COOJA.

The propagation model used in the simulations is COOJA's Unit Disk Graph Medium (UDGM) with constant loss. 
This model uses two concentric disks of different radius. The first radius is the transmission range while the second, larger radius is the interference range. Any node within the transmission range receives transmitted packets with a probability of 100\% (default value). Any node within the interference range but outside of the transmission range is not able to receive any packets but see its own transmissions affected. Finally nodes outside of the interference range are not able to receive packets and are not affected by transmitted packets.
%This propagation model considers that a frame is propagated to all nodes within a circle around the transmitter. The radius of this circle is the transmission range. In this range we have kept the 100\% default values for both transmission and reception success probabilities. Moreover, a node within the larger interference range, can't receive the packet but is subject to collisions caused by the transmission.

% -----------------------------------------------------------
\subsection{Configuration of protocols}

The default parameters of ContikiMAC and X-MAC have been used. All
protocols use a wake-up interval of 125ms (default value within
Contiki). X-MAC uses the default 5\% duty cycle (as specified in
Contiki). Due to the \emph{fast sleep optimization} the duty-cycle of
ContikiMAC is ``dynamic" and therefore cannot be
specified. Table~\ref{tab:cfg} summarizes the different parameters
used.

The configuration of both protocols has also been modified as
follows. The number of CCAs used by the ContikiMAC collision
avoidance mechanism has been changed from 6 to 2. This is enough to
satisfy the constraint $T_i < T_c$. Using more than two
CCAs would result in an increase of latency without any benefits.
   
 %T_i=0.4 (contiki code) T_c=0.5 (contiki code)  
% For X-MAC, the broadcast mechanism uses the ``strobe based" approach instead of the default ``data repetition" approach (cf Section~\ref{sec:background1}). The aim is to focus on the different transmission procedures between X-MAC and ContikiMAC.
   
The X-MAC \emph{streaming} mode and its counterpart in ContikiMAC,
called \emph{burst} mode, have both been disabled. This allows to
focus only on standard transmissions and ignore the cases where the
destination is already awake after the reception of a previous packet
in the same data flow.

When X-MAC or ContikiMAC fail to transmit a frame, the above layer will
schedule a retransmission. Failure to transmit occurs on a lack of
strobe-ACK (X-MAC) or lack of data-ACK (X-MAC and ContikiMAC). The
number of retransmissions is limited to 3. Note that in Contiki,
retransmissions are performed by the CSMA module.

\begin{table}[ht!]
  \begin{center}
    \begin{tabular}{|r|c|c|}
      \cline{2-3}
      \multicolumn{1}{c|}{} & \textbf{X-MAC} & \textbf{ContikiMAC} \\
      \hline
      Wake-up interval & 125ms & 125ms \\
      \hline
      Duty-cycle & 5 \% & dynamic \\
      \hline
      \# of CCAs & 6 & 2 \\
      \hline
%      $T_i$ & N/A & 0.4 ms \\
%      \hline
%      $T_c$ & N/A & 0.5 ms \\
%      \hline
      Streaming / burst mode & disabled & disabled \\
      \hline
      \# of \textbf{re}transmissions & 3 & 3 \\
      \hline
      Data packet size & 87 bytes & 87 bytes \\
      \hline
    \end{tabular}
  \end{center}\caption{Summary of the parameters used for our experiments.}\label{tab:cfg}
\end{table}

%6CCA are needed to cover two strobes
 
 % -----------------------------------------------------------
\subsection{Network topologies}

We consider two different kinds of scenarios to compare the
protocols. First, we use small star topologies with static
routing (see Section~\ref{sec:topologies-static}). The size and
traffic of those topologies can be scaled arbitrarily and their
density controlled. Moreover, as no routing protocol is running, the
only traffic that will be exchanged between the nodes is the ``application'' traffic
we inject.

Second, we use larger more realistic topologies where a collect-type
application is in use (see Section~\ref{sec:topologies-routing}). The
routes towards the sink are computed using the RPL
protocol~\cite{rpl-rfc}.

% - - - - - - - - - - - - - - - - - - - - - - - -
\subsubsection{Star topologies --- static routing}
\label{sec:topologies-static}

The star topologies consist in one central node surrounded by $N$ neighbours where $N$ varies from 2 to 14. Those $N$ neighbours are organized in $N/2$ (\textit{source}, \textit{destination}) pairs. Each sender is only allowed to send data to its assigned destination. Node 1 is a relay. The nodes with an even ID are senders. Pairs are formed as follows: a sender with the ID $k$ is assigned a destination at the opposite side of the circle. For example with $N$=14, the following pairs are formed: (2,9), (4,11), (6,13), (8,15), (10,3), (12,5) and (14,7).

\begin{figure}[ht!]
  \begin{center}
    \includegraphics[scale=0.45]{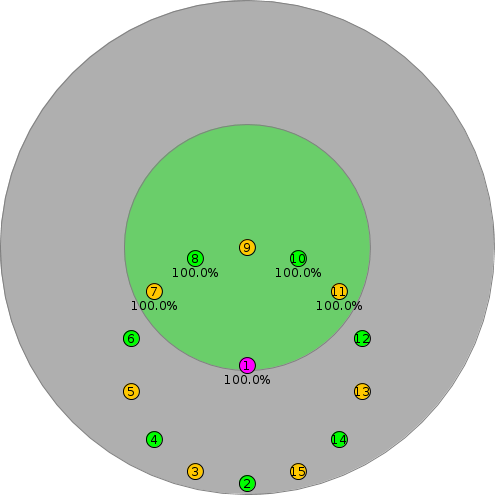}
    \caption{Density/bandwidth test topology with 14 neighbours. The effective transmission and interference ranges of node 9 are respectively represented by the green and grey disks.}
    \label{fig:densityTopology}
  \end{center}
\end{figure}

Fig.~\ref{fig:densityTopology} illustrates such a topology. The green nodes depict the sources while the yellow nodes the destinations. In addition, the green disk around node 9 represents its transmission range while the grey disk represents its interference range.

Routes are computed in advance so that each sender node has a default route through the central node. As a consequence, every network-layer communication between a source $S$ and a destination $D$ requires 2 link-layer transmissions: first from $S$ to the central node and second from the central node to $D$.

The test topologies are evaluated with different transmission rates. Each source node performs an application-level data packet transmission towards its corresponding destination at a regular interval. The lower this interval, the higher the transmission rate. Three different rates ($R$) are used between transmissions :  \textit{high} (interval=5s), \textit{moderate} (interval=10s) and \textit{low} (interval=15s).

For each topology size ($N$) and rate ($R$), 25 runs of the experiment are performed. Each experiment lasts 4 simulated minutes.

% - - - - - - - - - - - - - - - - - - - - - - - -
\subsubsection{Realistic topologies --- RPL routing}
\label{sec:topologies-routing}

In a second step we perform simulations in a more realistic setup. For these scenarios the nodes are running a collect application where each selected transmitter node randomly sends packets to the sink with an interval of 15 seconds.The routes are computed dynamically using RPL~\cite{rpl-rfc}. 
The Contiki RPL configuration is used. Each experiment lasts 4 simulated minutes\footnote{In 90\% of all experiments, RPL converges in less than 35 seconds}. Fifty runs are executed each one with a different topology.

\begin{figure}[ht!]
  \begin{center}
    \includegraphics[scale=0.45]{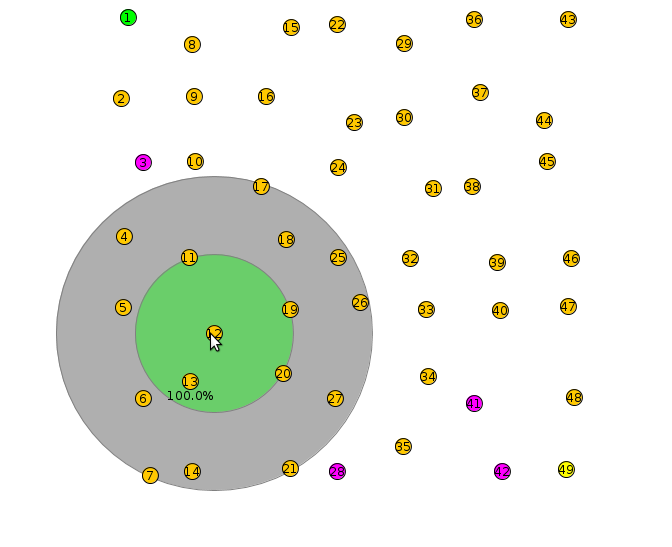}
    \caption{The 49-nodes topology. The green (resp. grey) disk represent the transmission (resp. interference) range for node 12.}
    \label{fig:netView}
  \end{center}
\end{figure}

The topologies are generated randomly. Nodes are positioned on a square grid, their coordinate being picked randomly according to uniform random variates. Each topology contains 49 nodes, 48 of them being sources and the remaining one being the sink. Fig.~\ref{fig:netView} gives an example of one of the topologies used. The sink is always the node at the top left-corner. 

Traffic is generated as follows. For each simulation 5 active nodes are randomly selected and the 44 remaining nodes stay passive. A passive node doesn't send any application layer packet but is allowed to forward the traffic originated by active nodes.

Each MAC protocol allows a maximum of 4 transmission attempts before dropping the packet.

% -----------------------------------------------------------
\subsection{Metrics under consideration}
\label{sec:metrics}

We consider the following metrics: the latency, the expected
transmission count (ETX), the number of retransmissions
%(due to a collision or lack of acknowledgement)
, the packet delivery ratio (PDR) and the duty cycle. We detail those
metrics in the following paragraphs.

The \textbf{latency} is measured as the average end-to-end
transmission time, that is the time needed by a sender to transmit a
packet to a destination potentially across multiple hops. 
The latency includes the retransmissions needed in case of collision or lack of
acknowledgement.

The \textbf{Expected Transmission Count (ETX)}, is the expected number
of hop-to-hop transmissions needed by a packet to be received without
error by the destination. We use this metric to estimate the average number of
retransmissions required by hop-to-hop transmissions.

\begin{figure}[ht!]
  \begin{center}
    \includegraphics[scale=0.4]{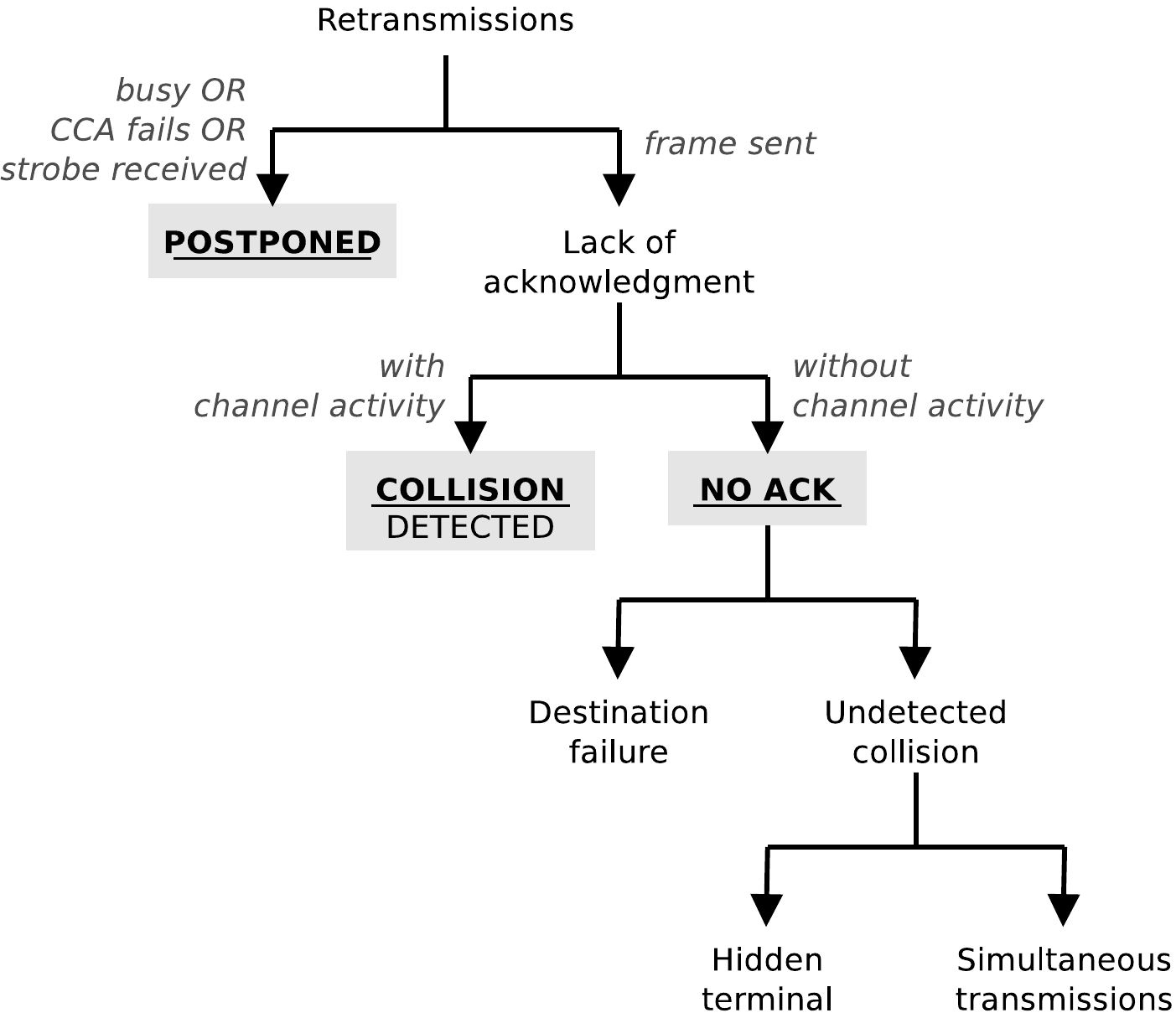}
    \caption{Classification of retransmission causes.}
    \label{fig:retx-causes}
  \end{center}
\end{figure}

 We count a
retransmission in the following cases
(Fig.~\ref{fig:retx-causes}). 
First, the sender can \textbf{postpone} a transmission in the
following cases: (1) due to an incoming transmission, a packet is already pending
in the radio, (2) the channel is stated as busy by the CCA collision
avoidance mechanism (\implem{ContikiMAC}, \implem{X-MAC-C} and
\implem{X-MAC-CP}) or (3) a strobe has been detected recently
(\implem{X-MAC} and \implem{X-MAC-P}).
Second, the transmission can \textbf{collide} with another ongoing
transmission. This case occurs when no ACK is received but channel
activity is detected instead.
Third, \textbf{no ACK} is received after the transmission and no
channel activity is detected. This can occur due to one of the
following reasons : (a) the destination node has failed or is out of
range, (b) a collision occurred at the receiver but is not detected by
the sender (hidden terminal) or (c) another sender started its
transmission simultaneously, making the collision avoidance mechanism
ineffective.

%  \begin{figure}[ht!]
%   \begin{center}
%     \includegraphics[scale=0.6]{pictures/concurrent.png}
%     \caption{In case of two simultaneous transmissions, the CCA collision avoidance mechanism can be powerless.}
%     \label{fig:concurrent}
%   \end{center}
% \end{figure}

The \textbf{PDR} is measured as the ratio between the number of
packets sent to a destination and the number of packets received by
this destination.

The \textbf{Duty Cycle} provides information regarding the energy
consumption of a node by evaluating the time spent by a node in the following
states : \textit{listen}, \textit{RX}, \textit{TX}. The Duty Cycle is expressed as the ratio between the time spent by a node in those three states and the wake-up interval. The Duty Cycle value is computed via the Cooja \textit{Powertracker} tool.

% =====================================================================================
\section{Star topologies without routing}
\label{sec:results-regular-topo}

This section presents the results obtained with star topologies and
static routing. The results have been averaged on 25 runs.
The main lesson is that \implem{ContikiMAC} clearly outperforms
\implem{X-MAC} in terms of latency and energy consumption. However,
the results also show that some of our X-MAC variants can achieve
results close to that of \implem{ContikiMAC} and we explain why.

%-------------------------------------
\subsection{Expected number of transmissions}
\label{sec:results-regular-topo-retx}

\begin{figure}[h!]
\begin{center}
\subfloat{}{\includegraphics[scale=0.39]{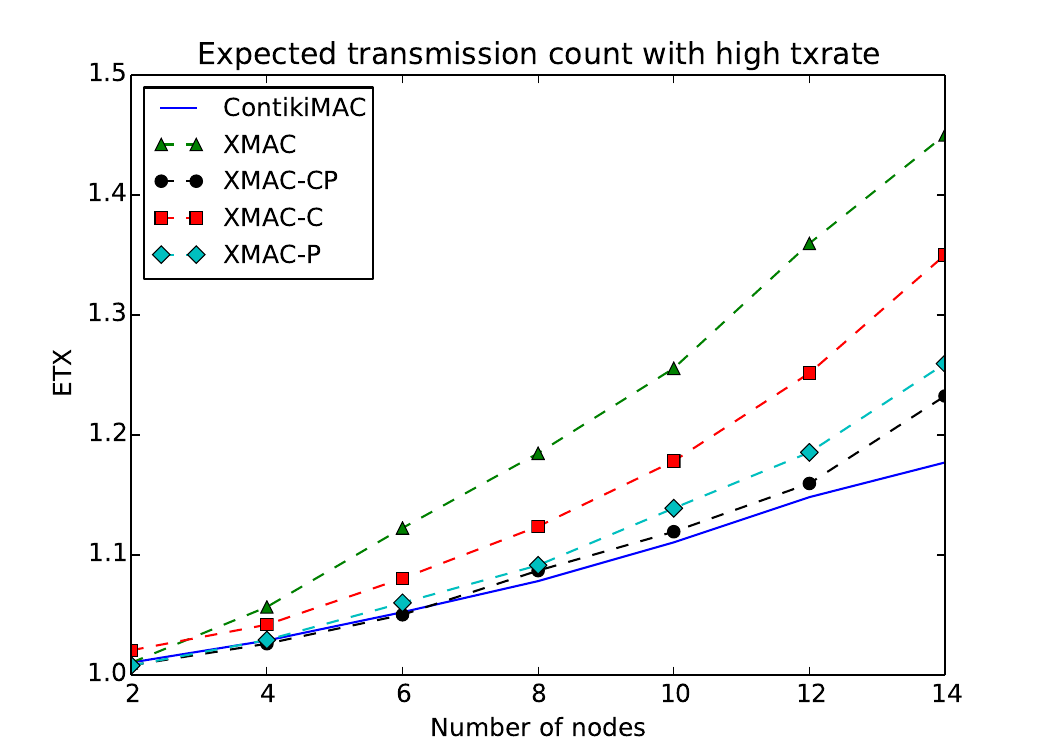}}
\subfloat{}{\includegraphics[scale=0.39]{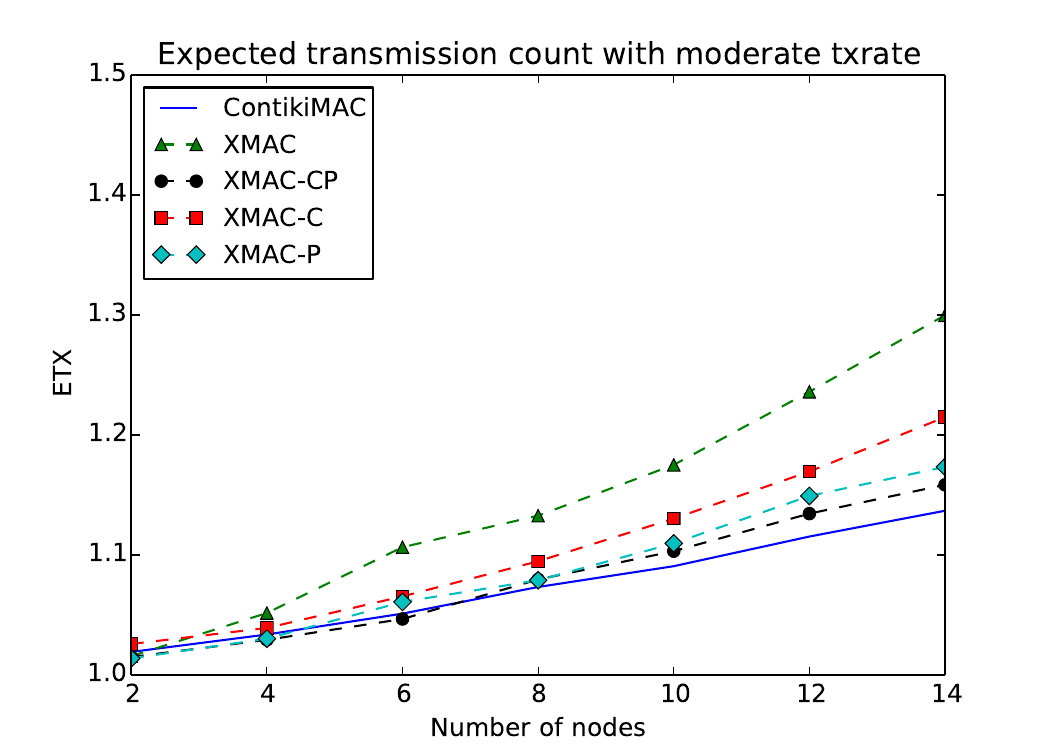}}
\subfloat{}{\includegraphics[scale=0.39]{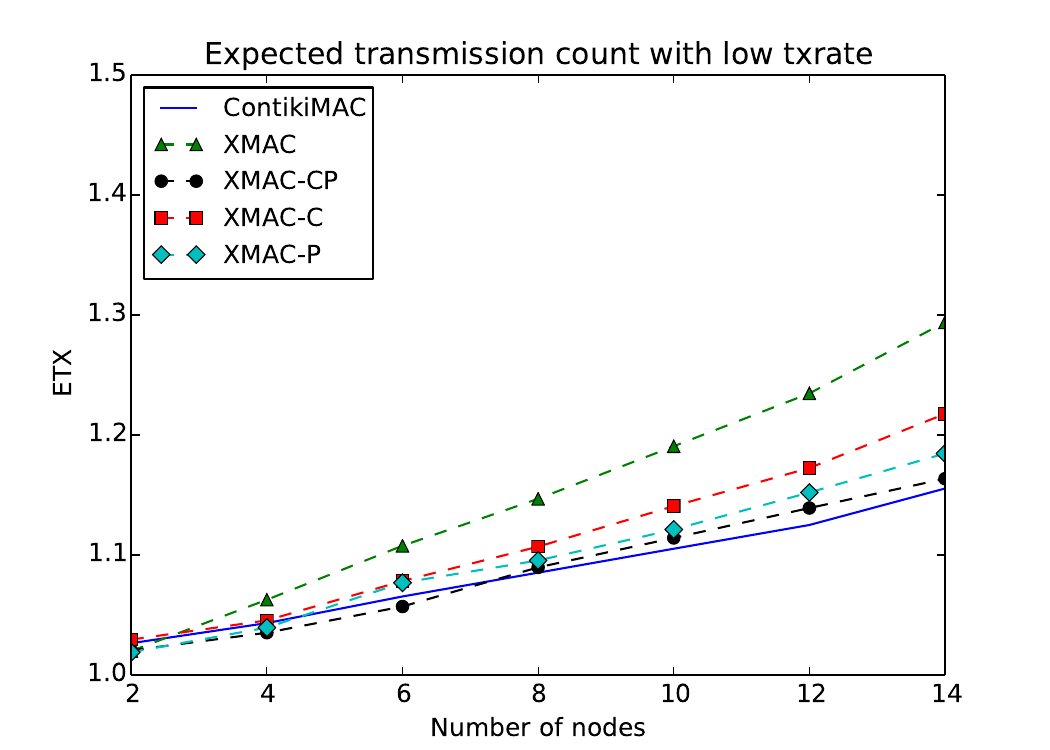}}
\caption{Average ETX.}
\label{fig:densityReTX}
\end{center}
\end{figure}

We first focus on the ETX as it has a significant impact on
latency and will be used later to explain latency
results. Fig.~\ref{fig:densityReTX} shows the average ETX as a function of the number of nodes. Each
subfigure corresponds to a different transmission rate (high,
moderate, low). We don't show the standard deviation as is it always
lower than 1\%.

Fig.~\ref{fig:densityReTX} shows that \implem{ContikiMAC}
 consistently requires the lowest number of transmissions attempts. On the
 opposite, \implem{X-MAC} has the highest ETX. 
 The X-MAC variants are in-between and their order remain the same when the number of nodes and the traffic intensity vary. The lowest ETX of ContikiMAC can be attributed to two causes : the \textbf{
   collision avoidance mechanism} and a lower channel use due
 to the \textbf{phaselock}. We discuss these two causes separately in the next paragraphs by looking at the
 results obtained with our modified X-MAC versions.

% \begin{figure}[ht!]
% \begin{center}
% \begin{minipage}[h!]{\linewidth}
% \includegraphics[scale=0.4]{pictures/density/new-density-ok-5.pdf}
% \end{minipage} \hfill
% \begin{minipage}[h!]{\linewidth}
% \includegraphics[scale=0.4]{pictures/density/new-density-ok-10.pdf}
% \end{minipage}
% \centering
% \begin{minipage}[h!]{\linewidth}
%  \includegraphics[scale=0.4]{pictures/density/new-density-ok-15.pdf}
% \end{minipage}
% \caption{Average number of transmissions.}
% \label{fig:densityReTX}
% \end{center}
% \end{figure}

First, as expected, the CCA-based collision avoidance mechanism
\uline{significantly reduces the number of collisions}. When a
collision occurs, the two nodes involved in the collision need to
retransmit. To the contrary, when a node postpones its transmission
due to another ongoing transmission, a single retransmission is
needed. This reduction can be observed by comparing the results of
\implem{X-MAC-C} with that of \implem{X-MAC}.
%The use of the CCA based collision avoidance mechanism
%``considerably'' decreases the number of retransmissions due to
%collisions, compared to \implem{X-MAC}.
For example, at the highest density and rate (top subfigure in
Fig.~\ref{fig:densityReTX}), the CCA mechanism decreases the average
number of transmissions from 1.45 down to 1.35. 

Second, the phase-lock mechanism \uline{significantly reduces the
  channel use} and, as a consequence, reduces the average number of
retransmissions. This can be observed by comparing the ETX of
\implem{X-MAC-C} and \implem{X-MAC-P} : the later requires a lower
number of retransmissions. At the highest rate and density, the ETX is
reduced from 1.35 with \implem{X-MAC-C} down to 1.26 with
\implem{X-MAC-P}.

Third, the best results are obtained by the joint use of phase-lock
and CCA-based collision avoidance. This allows \implem{X-MAC-CP} to
further reduce the ETX compared to \implem{X-MAC-P} and perform
similarly to \implem{ContikiMAC}. In the end, at the highest rate and
density, the ETX has been reduced from 1.45 with \implem{X-MAC} down
to 1.23 with \implem{X-MAC-CP}, slightly above \implem{ContikiMAC}
which achieves an ETX of 1.18.

Finally, it must be noted that \implem{ContikiMAC} and the X-MAC
protocols differ in the way they achieve transmissions. This
difference has an impact on the performance, especially at the highest
transmission rates. The original X-MAC protocol consider data-ACK as
optional (see Section~\ref{sub:sec: contikix-mac}). Therefore, even if data-ACK are used, X-MAC
considers a transmission successful as soon as the data frame has been
sent. However, in several situations, the data could not be delivered (cf Sec~\ref{sec:metrics}). In this case no data ACK is received and the upper layers need to schedule a
re-transmission. \implem{ContikiMAC} acts differently as the repeated transmission of data packets is only interrupted once a data-ACK is received (or a collision detected).

\subsubsection{Causes of retransmissions}
  
To better understand the impact of each protocol on retransmissions and to confirm the above hypotheses we look deeper at the causes of
retransmissions. Table~\ref{tab:breakdown} provides for the highest
density and rate a breakdown of the retransmissions according to their
causes. The last 3 columns correspond to the 3 causes we identified in
Section~\ref{sec:metrics} : the transmission has been
\textbf{postponed} by the collision avoidance mechanism, a
\textbf{collision} occurred during the transmission, or \textbf{no
  ACK} was received. The columns provide the average occurrence ratio
of each cause.

\begin{table}[ht!]
\small
\begin{center}
    \begin{tabular}{|l|l|l|l|l|}
      \hline %computed as Postpone/(Postpone+Collision+noACK)
       \textbf{Protocol} & \textbf{ReTX (\#)} & \textbf{Postpone (\%)} & \textbf{Collision (\%)} & \textbf{no ACK (\%)} \\
      \hline
      \implem{ContikiMAC} & 148 & 44.05 & 10.7 & 45.25\\
      \hline
      \implem{X-MAC} & 334 & 13.9 & 54.5  & 31.6\\
      \hline
      \implem{X-MAC-C} & 255 & 78.6 & 4.8 & 16.6 \\
      \hline
      \implem{X-MAC-P} & 190 & 11.5 & 66.3 & 22.2 \\
      \hline
      \implem{X-MAC-CP} & 172 & 60.1 & 22.1 & 17.8\\
      \hline
    \end{tabular}
    \vspace{0.5cm}
    \caption{Breakdown of the ReTX causes with high transmission rate and density}
    \label{tab:breakdown}
    \end{center}
\end{table} 

If we focus on the results of \implem{X-MAC} and \implem{X-MAC-C} in
Table~\ref{tab:breakdown}, we observe that the fraction of \textit{postpone} transmissions drastically increases. This is due to the CCA collision avoidance
mechanism which allow to postpone a lot of transmissions which would
otherwise have caused collisions : more than 78\% of the
retransmissions in \implem{X-MAC-C} are postponed transmissions while
this amounts to only about 14\% in \implem{X-MAC}. This confirms the
efficiency of the CCA-based collision avoidance in reducing the number
of retransmissions.

The phaselock in \implem{ContikiMAC}, \implem{X-MAC-P} and
\implem{X-MAC-CP} reduces the number of retransmissions compared to
\implem{X-MAC} and \implem{X-MAC-C}. The least reduction is obtained
with \implem{X-MAC-P}. The reduction
obtained by \implem{X-MAC-P} compared to \implem{X-MAC-C} is
mainly due to reduced channel use. However, \implem{X-MAC-P} being free of CCA mechanism,  the
fraction of retransmissions due to collisions remains very high
(more than 66\%). As already observed above, the combination of phaselock
and collision avoidance provides the best results : \implem{X-MAC-CP}
and \implem{ContikiMAC} have the least retransmission counts, a large
fraction of these being postponed transmissions.

The highest fraction of retransmissions caused by \textbf{no ACK} for
\implem{ContikiMAC} (~45\%) versus \implem{X-MAC-CP} (~18\%) is
explained by the fact that after the reception of a strobe-ACK
\implem{X-MAC-CP} performs a CCA to ensure that the medium is still
free before sending the data frame. This gives \implem{X-MAC-CP} an
additional opportunity to prevent collisions.

%involves more retransmissions due to collision or
%lack of ACK. This is explained by a reduced effectiveness of the CCA
%collision mechanism due to the fact that the nodes, targeting the same
%phaselock time, will start their CCA evaluation at the same
%time. \NDLR{Verify link with above hypotheses.}

%-------------------------------------
\subsection{Latency}
\label{sec:results-regular-topo-latency}

\begin{figure}[h!]
\begin{center}
\subfloat{}{\includegraphics[scale=0.39]{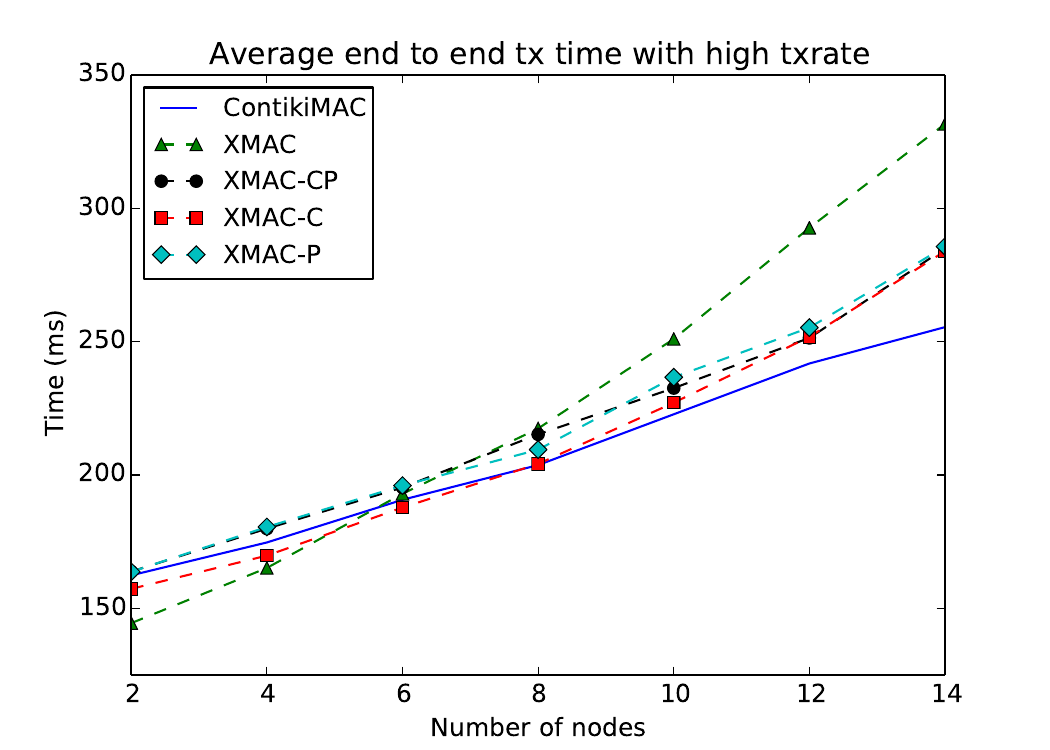}}

\subfloat{}{\includegraphics[scale=0.39]{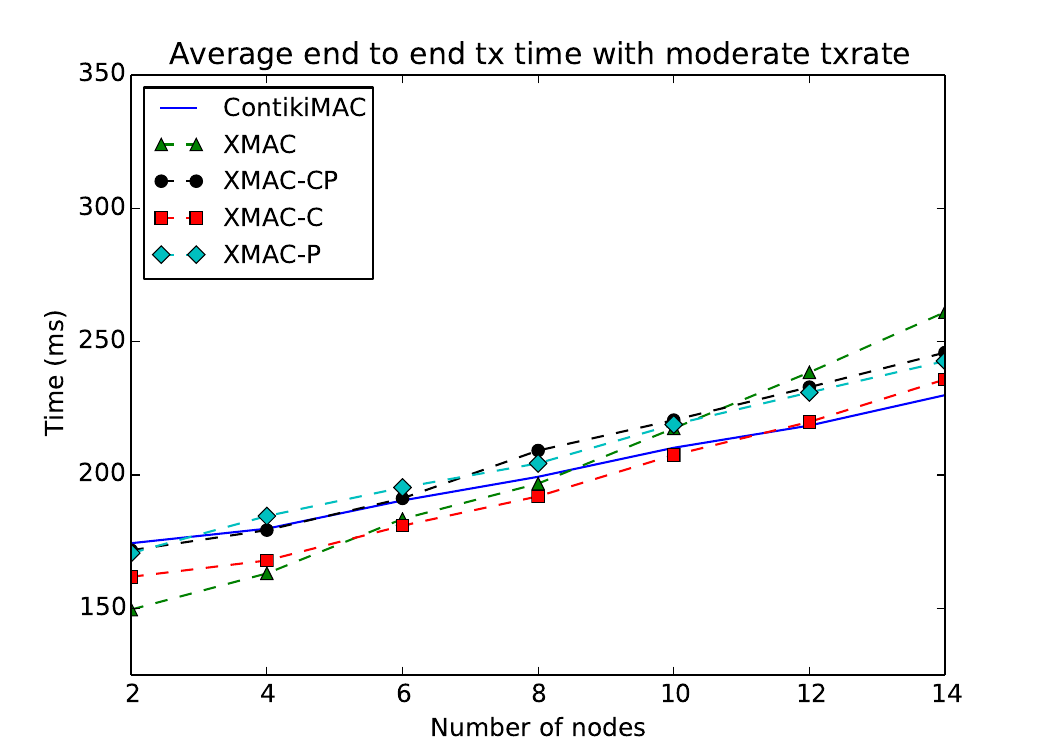}}

\subfloat{}{\includegraphics[scale=0.39]{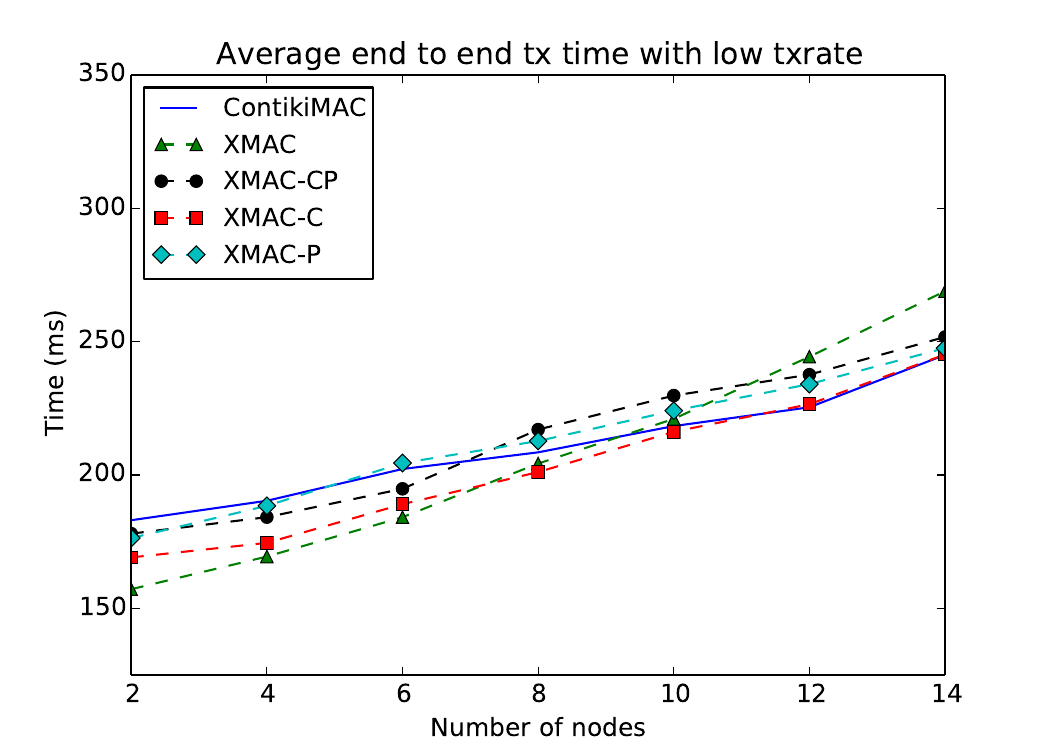}}
\caption{Average end-to-end transmission time.}
\label{fig:densityTxTime}
\end{center}
\end{figure}

This section compares the end-to-end latency of the different MAC
protocols. The results are shown in Fig.~\ref{fig:densityTxTime} where
latency is given as a function of density. Each subfigure corresponds
to a different transmission rate (high, moderate, low).

A first look at the results leads to the following
observations. First, at low densities protocols without phaselock
(\implem{X-MAC} and \implem{X-MAC-C}) achieve lower latencies than
protocols with phaselock (\implem{X-MAC-P}, \implem{X-MAC-CP} and
\implem{ContikiMAC}). Second, the latency increases with density and
at some point phaselock protocols start to perform better and achieve
the lowest latencies. Third, regardless of the traffic rate, at the
highest densities, \implem{ContikiMAC} always achieves the lowest
latency. We explain these three results in the following paragraphs.

\subsubsection{Phaselock at low densities}
At low densities, the phaselock seems detrimental to
latency. Fig.~\ref{fig:densityTxTime} shows that \implem{X-MAC} and
\implem{X-MAC-C} achieve lower latencies than protocols with phaselock
(\implem{X-MAC-P}, \implem{X-MAC-CP} and \implem{ContikiMAC}). As an
example, the average latency of \implem{X-MAC} at the highest rate and
lowest density is 144.4 ms  while it is 162.3 ms for \implem{ContikiMAC}.

This result is explained by a lack of accuracy in the phaselock
mechanism. The phaselock mechanism relies on the reception time of a
strobe-ACK (X-MAC variants) or data-ACK (ContikiMAC) to estimate the
receiver wake-up time. However, this method cannot learn the exact
wake-up time. The estimated wake-up time will often be slightly offset
(later) than the actual wake-up time and this causes an additional
delay in the end-to-end transmission. We observed that the offset
varies between 1 ms and 7.5 ms for \implem{ContikiMAC} while it can go
as high as 11.5 ms with \implem{X-MAC-P} and \implem{C-MAC-CP}. The
shorter offsets of ContikiMAC are explained by its fast-sleep
optimization mechanism. As a ContikiMAC receiver spends less time
idle, it is far less likely that it will answer at a time distant from its wake-up.% the receiver will wake-up earlier
than the estimated wake-up time.
%x-mac: 6.15 (listen time - 0.1 of a strobe)+0.7+3.9+0.7  / c-mac : 3
%(listen time) - 3.5(1st frame) +0.9

\subsubsection{Phaselock at higher densities}
When the density increases, the latency achieved increases regardless
of the protocol used. This is explained by the increase of the number
of retransmissions with density (see
Section~\ref{sec:results-regular-topo-retx}). However, the increase in
latency is slower for \implem{X-MAC-C} and for phaselock protocols. At higher densities and
rates, \implem{X-MAC-P}, \implem{X-MAC-CP} and \implem{ContikiMAC} now
achieve smaller latency than \implem{X-MAC}. This is explained by the fact that the
phaselock reduces channel use, hence the probability of
retransmission. At some point the lower number of retransmissions
compensates the inaccurate phaselock. At the highest rate and density,
\implem{ContikiMAC} achieves an average latency of 255.4 ms while \implem{X-MAC} requires 331.6 ms.

\subsubsection{Full data frames win over strobes}

We observed earlier that, regardless of the transmission rate, at the
highest densities, \implem{ContikiMAC} achieves the lowest
latency. This is really where the full data frame transmission
procedure of \implem{ContikiMAC} makes a difference. Indeed,
\implem{ContikiMAC} is able to send a packet using only two data
frames when the wake-up time of a neighbor has been estimated by the
phaselock and there is no collision. In the same conditions,
\implem{X-MAC-CP} requires two strobes and one data frame.

Let's compute the theoretical minimum time required to perform a
single-hop transmission with these protocols. We do this under the
assumptions that no retransmission is needed and the neighbor wake-up
time has been accurately estimated. For \implem{ContikiMAC}, this time
is equal to

%empiric measure via observation on timeline

$$T_\implem{ContikiMAC} = T_{data} + T_i + T_{data} = 6.7 ms\qquad (1)$$     

where $T_{data}$ is the transmission time of a data frame (2.9 ms) and
$T_i$ is the interval between two data frames (0.9 ms). For
\implem{X-MAC-CP}, this time is equal to

$$T_\implem{X-MAC-CP} = T_{strobe} + T_{sInt} + T_{strobe} + T_{bfDat} + T_{data} = 9.1
ms\qquad (2)$$

where $T_{strobe}$ is the transmission time of a strobe (0.7 ms), $T_{sInt}$
is the interval between two strobes (3.9 ms) and $T_{bfDat}$ is the
interval between the first acknowledged strobe and the data frame (0.9
ms). Note that these numbers assume the Contiki X-MAC default settings. $T_{sInt}$ is computed
dynamically by X-MAC depending on the duty cycle (5\%).

Now let's use these single-hop transmission times to get an idea of
the end-to-end transmission times. At
each hop, the transmitting node needs to wait until the receiving node
is awake. Given a wake-up interval of 125 ms, and assuming the wake-up
schedules are uniformly distributed, the average waiting time
$T_{avgWait}$ is $62.5 ms$. With $N$ hops, the average end-to-end
transmission time (without retransmission) is given by

$$T_{avgE2E} = N \times (T_{avgWait} + T_{tx})\qquad (3)$$

where $T_{tx}$ is the single-hop transmission time (either
$T_\implem{ContikiMAC}$ or $T_\implem{X-MAC-CP})$. In this particular
setup (star topologies), we consider only two-hop transmissions. This
leads to the following \textit{theoretical} average minimal times.
  
\begin{enumerate}
\item \implem{ContikiMAC} : $2 \times (62.5 ms + 6.7 ms) = 138.4 ms$
\item \implem{X-MAC-CP} : $2 \times (62.5 ms + 9.1 ms) = 143.2 ms$
\end{enumerate}

  We can observe that those times are coherent with the absolute
  minimums shown in Fig.~\ref{fig:densityTxTime}. The longer delays are
due to channel sharing which causes growing latency with increasing
density.

\subsubsection{Collision vs lack of acknowledgement}
Despite a lower number of retransmissions it looks like
\implem{X-MAC-P} has a higher latency than \implem{X-MAC-C}. This is
explained by the fact that in case of collision detected (and avoided)
the Contiki CSMA implementation plans a retransmission at the next
wake-up schedule, while the lack of ACK involves a delay proportional
to a cycle length using a linear back-off so that the interval between
the transmissions increases with each retransmit. As illustrated in
Table~\ref{tab:breakdown} , while \implem{X-MAC-C} avoids a lot of
collisions, \implem{X-MAC-P} is more impacted by a lack of ACK. At
higher density and transmission rate, \implem{X-MAC-P} misses more
acknowledgements than \implem{X-MAC-C}.

It can also be observed that at lower density and bandwidth the basic
\implem{X-MAC} implementation has a lower latency than
\implem{X-MAC-C}. This is explained by the lack of CCA avoidance
mechanism which slightly delays the transmission to perform the
CCAs. The lower number of retransmissions doesn't compensate this
delay at low density.

\subsection{Packet Delivery Ratio}
\label{sec:results-regular-topo-pdr}
\begin{figure}[h!]
\begin{center}
\subfloat{}{\includegraphics[scale=0.39]{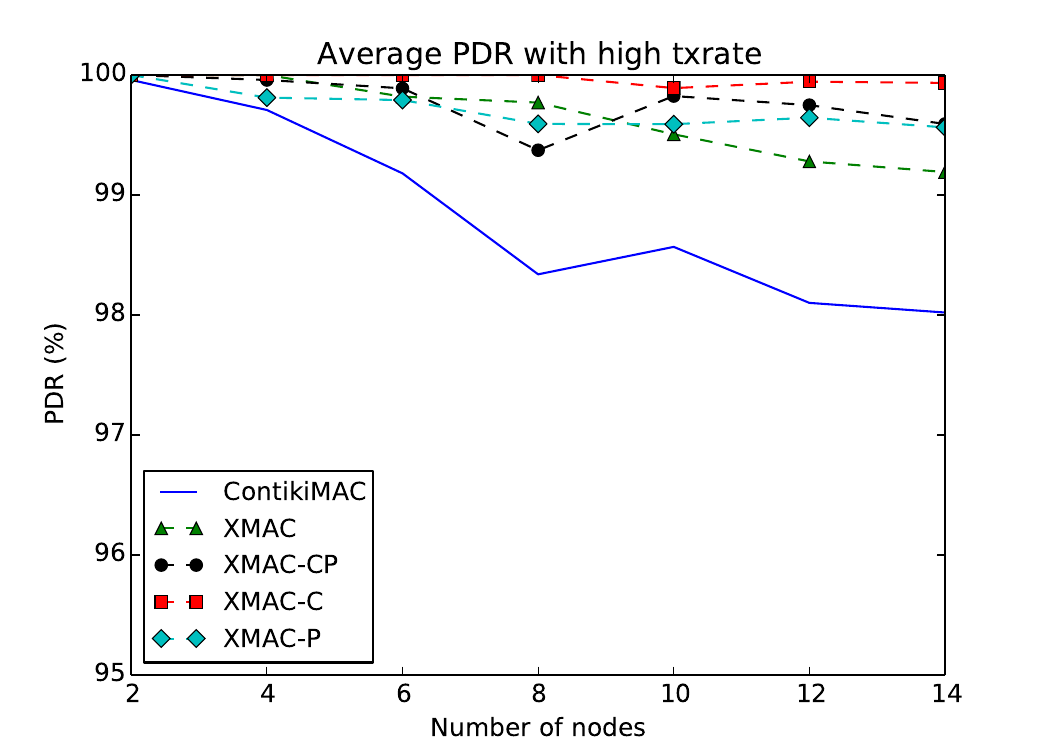}}
\subfloat{}{\includegraphics[scale=0.39]{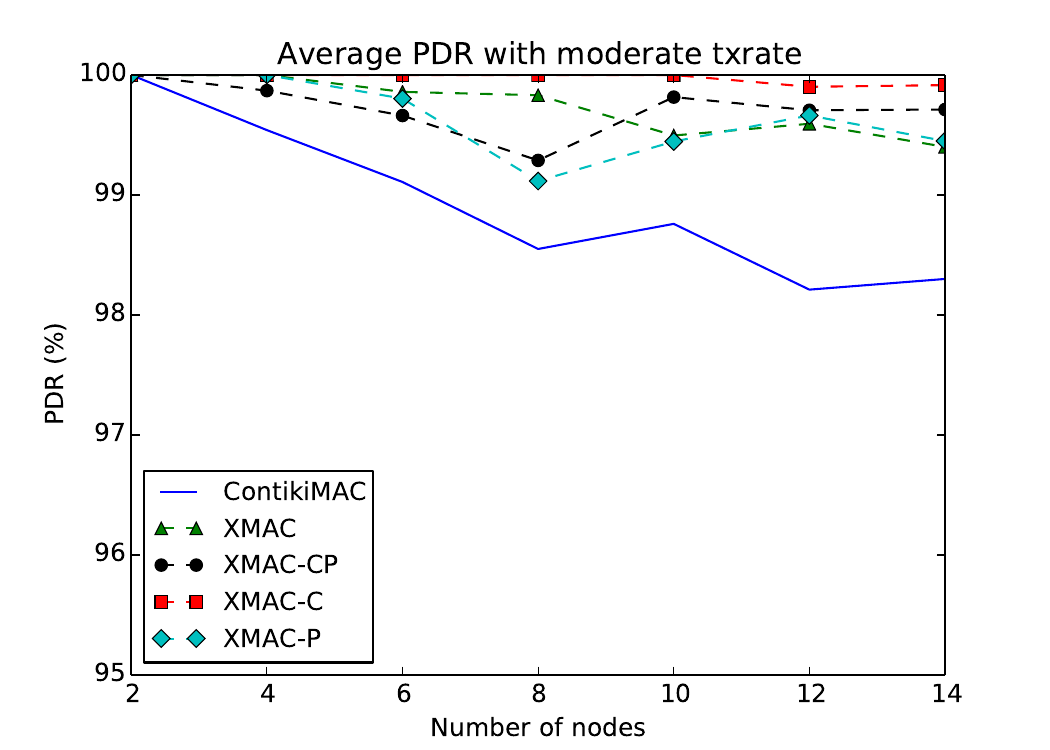}}
\subfloat{}{\includegraphics[scale=0.39]{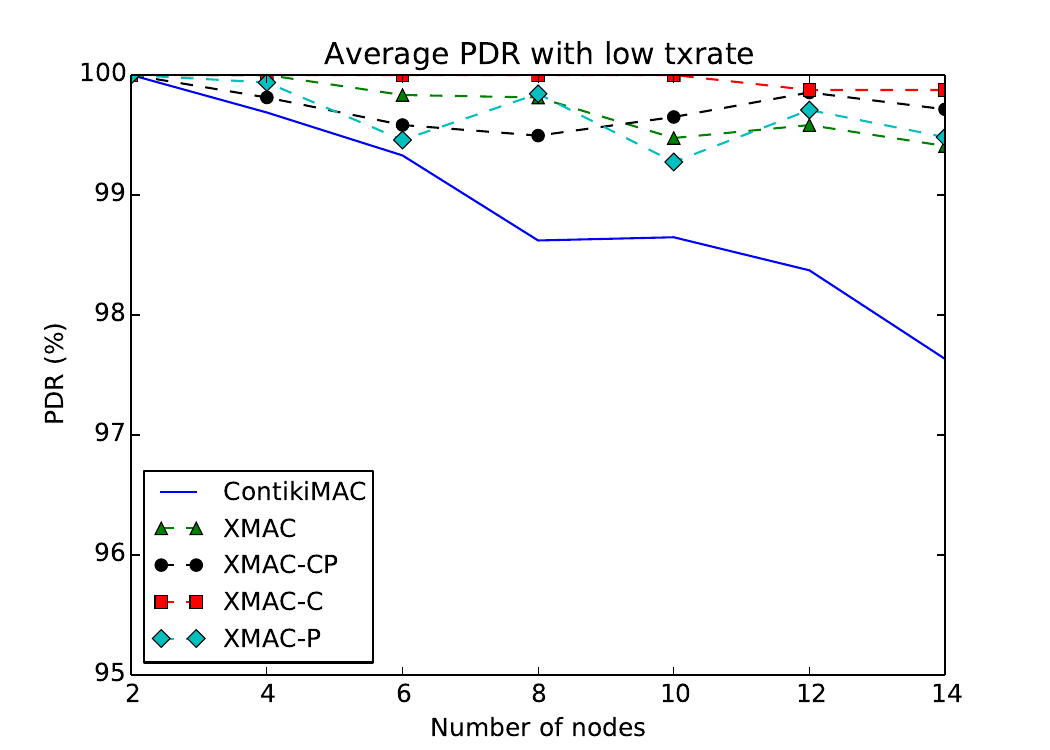}}
\caption{Average packet delivery ratio.}
\label{fig:densityPDR}
\end{center}
\end{figure}

In this section we study how the different MAC protocols behave regarding the Packet Delivery Ratio.
Fig.~\ref{fig:densityPDR} shows the packet delivery ratio obtained
with the different MAC protocols during our experiments. A striking
result is that \implem{X-MAC-C} consistently provides the highest PDR
(between 99 and 100 \%) while \implem{ContikiMAC} achieves the lowest
PDR with some values between 97 and 98 \%.

\begin{figure}[h!]
  \begin{center}
    \includegraphics[scale=0.38]{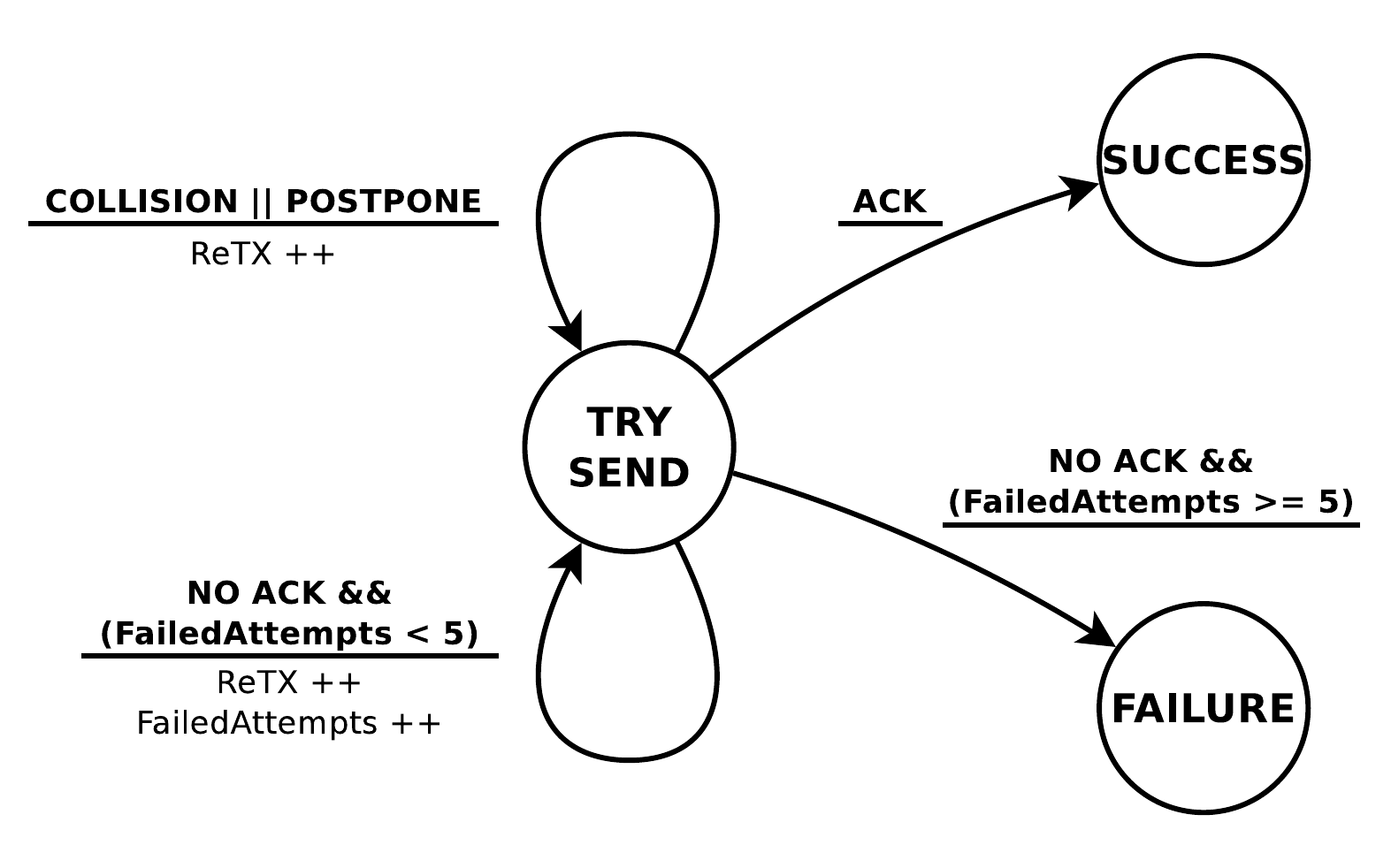}
    \caption{Handling of retransmissions in Contiki CSMA.}
    \label{fig:csma}
  \end{center}
\end{figure}

This difference can be explained by the way retransmissions are
handled. When the RDC layer (X-MAC or ContikiMAC) aborts the transmission of a frame, the upper layer, CSMA, can
decide to perform additional attempts. If the RDC layer postpones the transmission or detects a collision, a retransmission is always
performed by the CSMA layer. If on the other hand the transmission failed due to a lack
of acknowledgement (no ACK), up to 3 additional attempts are performed,
after which the transmission is aborted and reported as a
failure. Fig.~\ref{fig:csma} summarizes the behaviour of the CSMA layer and details how retransmissions (ReTX) are counted.

This way of handling retransmissions slightly handicaps
\implem{ContikiMAC}. As shown in Table~\ref{tab:breakdown}, the main cause of retransmission of \implem{ContikiMAC} is the
lack of acknowledgement (no ACK), which accounts for almost 47\%. To the opposite,
\implem{X-MAC-C} has a much lower number of such retransmissions :
about 16.5\% are due to no ACK. The number of frame transmissions
abandoned by CSMA is thus lower with \implem{X-MAC-C}, hence the higher PDR.

To confirm this explanation, we looked at the average percentage of frames which
were dropped by CSMA (after 4 attempts) during our experiments. The results for the highest density are
reported in Table~\ref{fig:csma-dropped-frames}. We observe that the
number of frames dropped by CSMA is much higher for
\implem{ContikiMAC} than for the X-MAC variants. \implem{X-MAC-C} has
the lowest number of frames dropped.
  
\begin{table}[ht!]
  \begin{center}
    \begin{tabular}{l|c|c|c|}
      \cline{2-4} & \multicolumn{3}{|c|}{\textbf{Frames dropped}} \\
      \hline
      \multicolumn{1}{|l|}{\textbf{Protocol}} & High txrate & Moderate txrate & Low txrate\\
      \hline
      \multicolumn{1}{|l|}{\implem{ContikiMAC}} & 2.05 \% & 1.8 \% & 2.52 \% \\
      \hline
      \multicolumn{1}{|l|}{\implem{X-MAC}} &  0.81 \% & 0.6 \% & 0.6 \% \\
      \hline
      \multicolumn{1}{|l|}{\implem{X-MAC-C}} &  0.06 \% & 0.08 \% & 0.13 \% \\
      \hline
      \multicolumn{1}{|l|}{\implem{X-MAC-P}} &  0.44 \% & 0.56 \% & 0.53 \% \\
      \hline
      \multicolumn{1}{|l|}{\implem{X-MAC-CP}} & 0.41 \%  & 0.29 \% & 0.29 \%\\
      \hline
    \end{tabular}
    \caption{Frames dropped by CSMA (highest density).}
    \label{fig:csma-dropped-frames}
  \end{center}
\end{table}

% \begin{table}[ht!]
%   \begin{center}
%     \begin{tabular}{|l|c|}
%       \hline
%       \multicolumn{1}{|c|}{\textbf{Protocol}} & \textbf{Frames dropped} \\
%       \hline
%       \implem{ContikiMAC} & 2.05 \% \\
%       \hline
%       \implem{X-MAC} &  0.81 \% \\
%       \hline
%       \implem{X-MAC-C} &  0.06 \% \\
%       \hline
%       \implem{X-MAC-P} &  0.44 \% \\
%       \hline
%       \implem{X-MAC-CP} & 0.41 \% \\
%       \hline
%     \end{tabular}
%     \caption{Frames dropped by CSMA (highest density and rate).}
%     \label{fig:csma-dropped-frames}
%   \end{center}
% \end{table}

Why does CSMA drops more frames when used with \implem{ContikiMAC}?
Because after a lack of ACK, \implem{ContikiMAC} will be mandated by CSMA to perform a retransmission. Thanks to its phaselock mechanism, \implem{ContikiMAC}
 will then re-schedule a transmission at the receiver's next wake-up.
The lack of acknowledgement can be due to an undetected collision. A node under
\implem{ContikiMAC} is more likely to reschedule the transmission at
the same time that the node responsible for the colliding
transmission. This leads to successive failures in row, until CSMA
drops eventually one of the colliding transmissions and explain why in
this configuration \implem{X-MAC} achieves a slightly higher PDR than
\implem{ContikiMAC}.

% In case of a lack of acknowledgement, \implem{ContikiMAC} will refer
% to its phaselock mechanism to try to schedule the retransmission at a
% time close to the destination's wake-up. Compared to \textbf{X-MAC},
% which is also impacted by a high lack of acknowledgement, this way to
% reschedule failed transmission is less random. If the lack of
% acknowledgement is due to an undetected collision, a node under
% \implem{ContikiMAC} is more likely to reschedule the transmission at
% the same time that the node responsible for the colliding
% transmission. This leads to successive failures in row, until CSMA
% drops eventually one of the colliding transmissions and explain why in
% this configuration \implem{X-MAC} achieves a slightly higher PDR than
% \implem{ContikiMAC}.

\emph{Although the phaselock mechanism reduces the number of retransmissions
for \implem{X-MAC-P}, \implem{X-MAC-CP} and \implem{ContikiMAC}, compared to \implem{X-MAC} and \implem{X-MAC-C}, it can 
sometimes negatively impact the PDR by synchronizing multiple transmitters.}
Consider for example the case of two transmitters that target the same receiver. This scenario is quite likely in our star topology where first hop transmission always target the central node.
%does not allow them to achieve a PDR as high as \implem{X-MAC-C}. The
%reason is there are still a significant number of collisions that
%cannot be avoided by performing a CCA. Indeed, the phaselock mechanism
%tends to synchronize nodes that want to transmit to the same
%receiver. 
As these nodes wake-up at almost the same time, they both
sense the channel as idle and start their transmission. In this setup
(star topologies), a first hop transmission always targets the
central node, which makes this event quite likely. Moreover, the
probability that two nodes wake-up at almost the same time increases
with the phaselock accuracy. This explains why \implem{X-MAC-CP}
performs better than \implem{ContikiMAC} in terms of PDR
since the phaselock offset is smaller with the latter. This can also
be confirmed by Table~\ref{tab:breakdown} where the fraction of
retransmissions due to a lack of ACK was almost 47\% with
\implem{ContikiMAC} while it was only 18\% with \implem{X-MAC-CP}.

% \begin{comment}
% It appears that \implem{X-MAC-C} is more efficient than the other protocols. The reason can be found on the way CSMA handle retransmissions: the CSMA mechanism drops a packet after five failed hop-to-hop transmissions. Obviously a packet dropped can't be received and due to that the PDR is linked to the number of retransmissions. However the Contiki CSMA implementation doesn't count an attempt due to a postpone triggered to avoid a collision.
% From Table~\ref{tab:breakdown} it appears that the postpone case is the main cause of collisions with \implem{X-MAC-C}.
% It explains why this protocol performs better than the other protocols. This is true even with the use of the phaselock mechanism which decreases the number of retransmissions.
% Indeed it appears that \implem{X-MAC-CP}, \implem{X-MAC-P} and \implem{ContikiMAC} achieve lower PDR than \implem{X-MAC}. This seems counter-intuitive as those three protocols have a lower number of retransmissions. This is explained by the fact that despite using phaselock, two nodes targeting the same destination will probably try to start their transmission at the same time. The collision avoidance mechanism will then be useless and the transmissions will collide avoiding the transmission of an ACK. This is confirmed by Table~\ref{tab:breakdown} where those protocols achieves a lower number of postpones than \implem{X-MAC-C}.
% \end{comment}

%-------------------------------------
\subsection{Energy consumption}
\begin{figure}[!h]
\begin{center}
\subfloat{}{\includegraphics[scale=0.39]{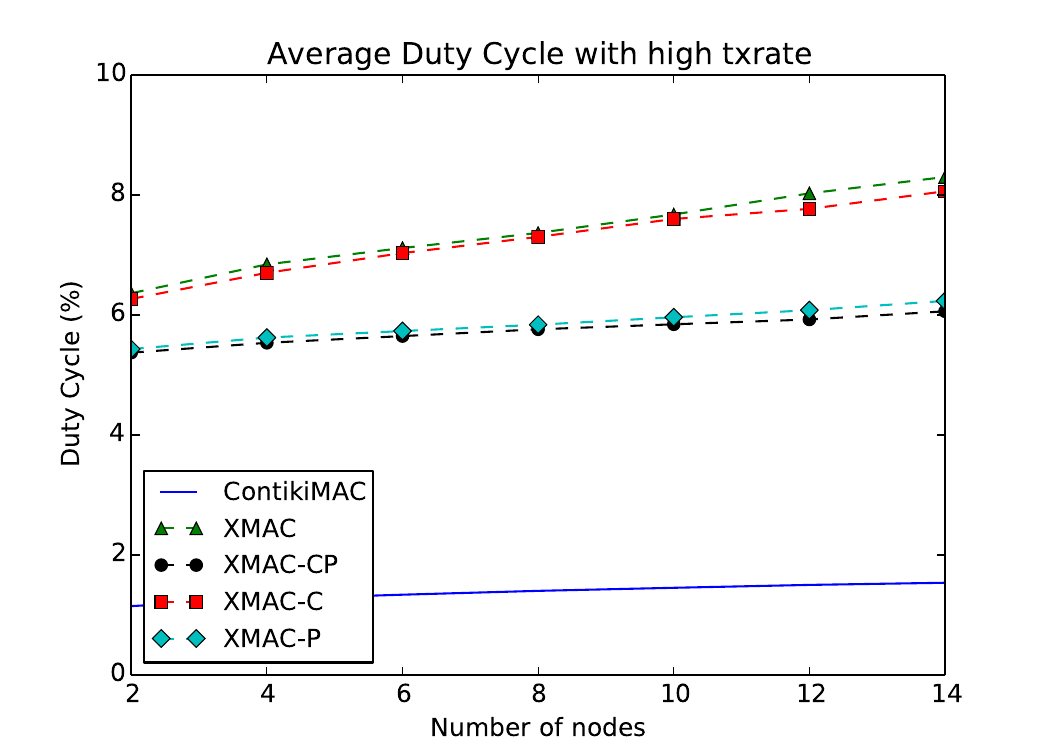}}
\subfloat{}{\includegraphics[scale=0.39]{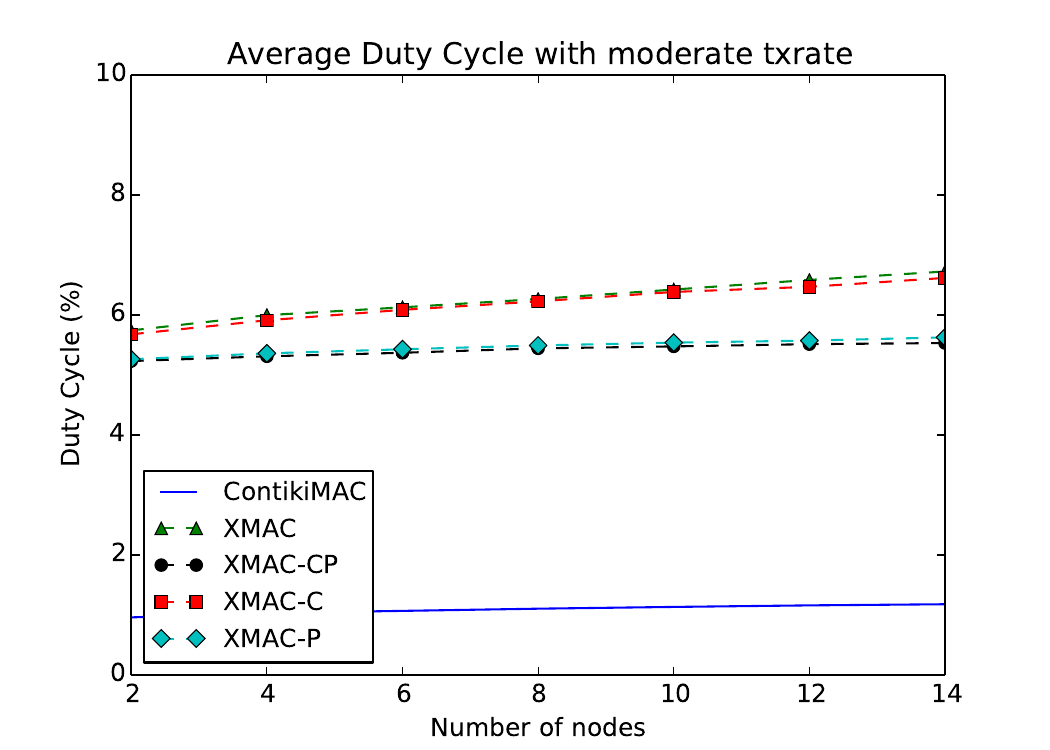}}
\subfloat{}{\includegraphics[scale=0.39]{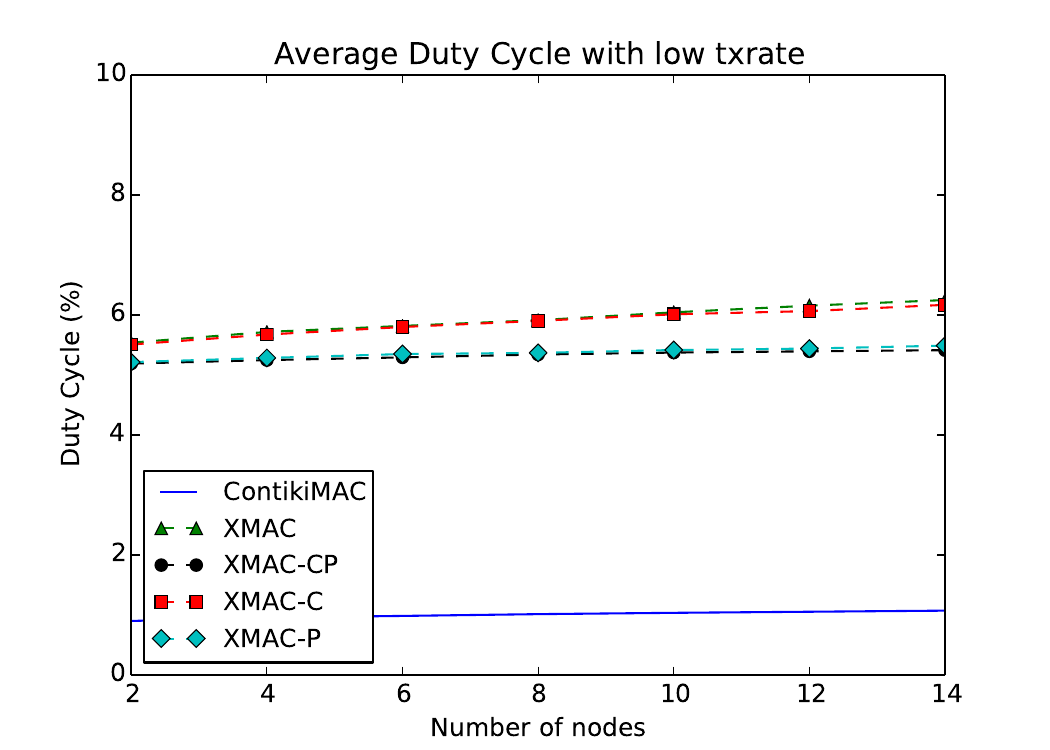}}
\end{center}
\caption{Average duty cycle.}
\label{fig:densityEnergy}
\end{figure}

%\label{sec:results-regular-topo-retx}
In this section, we look at the energy consumption of nodes under the different RDC protocols. 
We use the average duty-cycle as an \textbf{estimator} of the energy consumption. The
lower the duty-cycle, the longer the nodes batteries should last.
Fig.~\ref{fig:densityEnergy} represents the average network duty cycle ratio.
%for the time spent by each node in \textbf{\textit{listen}},
%\textbf{\textit{RX}} and \textbf{\textit{TX}} mode.% The duty-cycle
%provides a good indicator of the energy efficiency of a protocol. 

In our experiments, all the protocols are configured with the same
wake-up interval (125 ms). The main observation is that \implem{ContikiMAC} is
able to keep its duty-cycle \uline{6 times lower} than the most energy efficient X-MAC protocol (\implem{X-MAC-CP}).

To better understand what mechanism allows \implem{ContikiMAC} to reach that result, let's have a look at the results of the different variants.
 We can observe that the phaselock mechanism
plays an important role in the reduction of the duty-cycle by
comparing the results of the phaselock and non-phaselock variants of
X-MAC : the duty-cycle is reduced from 8\% down to 6\% at the highest
rate and density.
At the opposite, the introduction of the CCA-based collision
avoidance barely reduces the duty-cycle.\\ 
The most important contribution to the \implem{ContikiMAC}'s reduction in duty-cycle is the fast sleep
optimization at the heart of the wake-up procedure. This is confirmed by looking at the time spent by \implem{ContikiMAC} in \textit{listen} mode.
Tab.~\ref{tab:energy-breakdown} provides a breakdown of the time spent by the nodes in the following states: \textit{listen}, \textit{TX} and \textit{RX}. From this figures it appears that, with the highest density and transmission rate, \implem{ContikiMAC} spends almost five times less time in \textit{listen} state than \implem{X-MAC-CP}.

\begin{table}[h]
\footnotesize
\begin{center}
    \begin{tabular}{|l|l|l|l|}
      \hline
       & LISTEN (\%)\space\space & TX (\%)\space\space & RX (\%)\space\space\\
      \hline
      \textbf{ContikiMAC\space} & 1.21 & 0.21 & 0.10\\
      \hline
      \textbf{X-MAC} & 7.46 & 0.45 & 0.37\\
      \hline
      \textbf{X-MAC-CP} & 5.71 & 0.18 & 0.16\\
      \hline
      \textbf{X-MAC-C} & 7.26 & 0.43 & 0.36  \\
      \hline
      \textbf{X-MAC-P} & 5.84 & 0.20 & 0.17 \\
      \hline
    \end{tabular}
    \end{center}\caption{Average duty cycle statistics for highest density and txrate.}\label{tab:energy-breakdown}
\end{table}

Even if the duty-cycle for the four versions of X-MAC is fixed at 5\%,
we can observe higher values in the figures. This is due to the fact
that this duty-cycle can be extended by X-MAC if a node is involved in
any transmission: for example if the node detects a transmission at the end of its listening period,
 it will stay awake until the delivery of the packet, independently of its initial schedule.
% a broadcast transmission 

Given the duty cycle used by X-MAC defines the time spent in listening time, it would seem reasonable to 
reduce the prescribed duty-cycle of the X-MAC in order to reach energy consumption levels close to \implem{ContikiMAC}. For example, one might consider to use a duty cycle of 1\%.
But in fact this approach impacts negatively the other performance metrics. Indeed X-MAC needs on average two strobes to detect a
transmission. This implies that a node has to stay listening for at
least 5.3 ms after wake-up. This leads to a duty cycle ratio of at least 4.25\%.
 Reducing the duty cycle would cause a
decrease of the listening time and then the probability to detect
incoming traffic. We performed some experiments using a duty cycle of
1\% (similar to that of \implem{ContikiMAC}), but the PDR dropped
below the 5\% mark, due to the nodes being unable to detect incoming
frames.

% \begin{comment}
% Using the same 
% wake-up interval than \implem{X-MAC}, \implem{ContikiMAC} achieves a
% duty cycle 6 times lower. 
%  Even if the energy consumption of \implem{X-MAC-CP} is quite important compared to \implem{ContikiMAC}, the use of the phase-lock mechanism decreases that consumption compared to \implem{X-MAC} and \implem{X-MAC-C}. This consumption is still enhanced by the addition of the CCA-based collision avoidance mechanism.
% \end{comment}

%-------------------------------------
\subsection{Lessons learned}
\label{sec:results-regular-topo-lessons}
This section summarizes the main lessons learned from the simulations in star topologies with static routing. 
We observe first that ContikiMAC is both more energy and
latency efficient than any X-MAC variant. The addition of
phaselock and collision avoidance as done in \implem{X-MAC-CP} allowed
to achieve a latency similar to that of ContikiMAC. However, these
additions proved insufficient to reach a level of energy efficiency
equivalent to that of ContikiMAC. This was even more noticeable as
density and traffic rate increased.

For what concerns the PDR, \implem{X-MAC-C} provided the best success
rate at around 99-100\%. \implem{ContikiMAC} and \implem{X-MAC-CP}
performed slightly worse with a PDR between 97 and 98\% at the highest
density.  This result is explained by the phaselock mechanism synchronizing the transmitters which then face repeated collisions.
Despite this issue being partially an artefact of the star topology\footnote{Our star topology makes quite likely the case of two nodes sending a frame to the same
destination at the same time.}, investigating how to avoid collision when using a phaselock mechanism could be a potential lead for future works.

Within the number of retransmissions, a higher proportion of postponed transmissions combined with the
actual Contiki CSMA implementation explain the better PDR of
\implem{X-MAC-C}. In more realistic scenarios where most collisions
are due to ``hidden terminal'', we expect to see the difference
between ContikiMAC and X-MAC vanish. Given those collisions can't be detected, the number of
acknowledgements not received should increase while the number of
postpones decrease.

Finally, we have stated a lack of accuracy in the implementation of the phaselock mechanism. This can lead to errors of up several milliseconds when estimating the wake-up's phase of a node.

% -----------------------------------------------------------
\section{Realistic topologies - RPL routing}
\label{subsec:realScen}
\label{sec:results-random-topo}

This section presents the results of our experiments with larger,
randomly generated topologies and dynamic routing using RPL. The
results are summarized in Table~\ref{tab:metric1}. There is one row
per MAC protocol. The first column provides the average
packet delivery ratio (PDR). The second column provides the average
ETX along with the standard
deviation. The third column provides the end-to-end latency along with
the standard deviation among the different simulations.

\begin{table}[h]
\footnotesize
\begin{center}
    \begin{tabular}{|l|l|l|l|}% PDR = RX_DATA/TX_DATA
      \hline
      \textbf{} & PDR(\%) & ETX (\#) & Latency (s)\\
      \hline
      \textbf{ContikiMAC  }& \space95.3 & \space1.16 ($\pm$ 0.46)  & 1.07 ($\pm$ 0.80)\space\space\space \\
      \hline
      \textbf{X-MAC} & \space86.2 & \space1.54 ($\pm$ 1.18)  & 1.59 ($\pm$ 1.75) \\
      \hline
      \textbf{X-MAC-CP} & \space92.1  & \space1.28 ($\pm$ 0.71)  & 1.35 ($\pm$ 1.29)\\
      \hline
      \textbf{X-MAC-C} & \space92.3 & \space1.35($\pm$ 0.96)  & 1.25 ($\pm$ 1.29) \\
      \hline
      \textbf{X-MAC-P} & \space90.9 & \space1.27($\pm$ 0.74)  & 1.29 ($\pm$ 1.13) \\
      \hline
    \end{tabular}
    \end{center}\caption{Summary of experiments results.}\label{tab:metric1}
\end{table}

The following subsections discuss each column separately. A final
subsection discusses results in terms of duty-cycle and energy
consumption.

%--------------------------------
\subsection{Packet Delivery Ratio}
\label{sec:results-random-topo-pdr}

The PDR results are quite different from what we observed with star 
topologies. \implem{X-MAC} remains the weakest protocol with an
average PDR of 86.2\% while \implem{ContikiMAC} provides the highest
PDR (95.3\%). \implem{X-MAC-CP} has the closest results to
\implem{ContikiMAC}. We observe that \implem{X-MAC-C} is not the most
reliable protocol anymore, although it provides a PDR which is inline
with the other X-MAC variants. Overall, the PDR achieved by all the
protocols is much lower than that achieved with the star topologies.
This was expected given the larger average number of hops. In our RPL topologies, the average network diameter in our RPL is six hops as the average density.
 As single-hop success probabilities need to be multiplied together for all the links, this leads to an overall lower end-to-end success probability. Moreover, this experiment also includes RPL traffic in addition to the application traffic.
% \NDLR{How would you explain this. On one side, the average number of
%   hops is larger (how much), and as single-hop success probabilities need to be
%   multiplied together for all the links, this leads to an overall
%   lower end-to-end success probability. What about the traffic rate
%   and density ? How can we compare them with those of the star
%   topologies ?}

The difference between \implem{ContikiMAC} and the X-MAC variants is
explained by the lower reliability of X-MAC's transmission
procedure. X-MAC stops the transmission procedure as soon as it sends
the data packet after the reception of a strobe-ACK. If no data-ACK is
received, a brand new transmission has to been scheduled by CSMA. To the opposite, \implem{ContikiMAC}
waits for the data-ACK to consider the transmission done and stop the transmission procedure. 
Once a transmission rescheduled, the packet being not transmitted is then temporary stuck in the sender queues. This increases the use of the buffer and the probability than once the buffer filled a node can't handle an additional packet anymore. This why the X-MAC's transmission procedure facing more retransmissions can lead to lower PDR.

%--------------------------------
\subsection{Expected transmission count}
\label{sec:results-random-topo-retx}

The highest average expected transmission count is obtained with
\implem{X-MAC} (1.54) while the lowest is obtained with
\implem{ContikiMAC} (1.16). The collision avoidance and phaselock added in
the X-MAC variants play their role and the results are inline with
what we observed for the star topologies (see
Section~\ref{sec:results-regular-topo-retx}). The X-MAC's phaselock
variants require the least number of retransmissions : 1.28 and 1.27 for respectively \implem{X-MAC-CP} and \implem{X-MAC-P} against 1.35 for \implem{X-MAC-C}.

% \begin{comment}
% Based on the results observed in, the number of transmissions
% required to send successfully a single data packet (ETX) is
% considerably lowered by the use of the CCA collision avoidance and
% phaselock mechanisms. This explains the significantly smaller number
% of retransmissions of \implem{X-MAC-C} and \implem{X-MAC-CP} compared
% to the genuine \implem{X-MAC}. However \implem{ContikiMAC}, which
% achieved a better PDR than \implem{X-MAC-CP}, requires less
% retransmissions.
% \end{comment}

%--------------------------------
\subsection{Latency}
\label{sec:results-random-topo-latency}

As shown in Table~\ref{tab:metric1}, 
\implem{ContikiMAC} achieves the lowest average end-to-end latencies at
1.07 s. Moreover \implem{ContikiMAC}'s latency varies less among the different transmissions as shown by the standard
deviation. X-MAC protocols show higher latency
variability. 
This is coherent with the results in terms of retransmissions
discussed in Section~\ref{sec:results-random-topo-retx}.

\begin{figure}[h]
\begin{center}
\hspace{-25px}
   \includegraphics[scale=0.3]{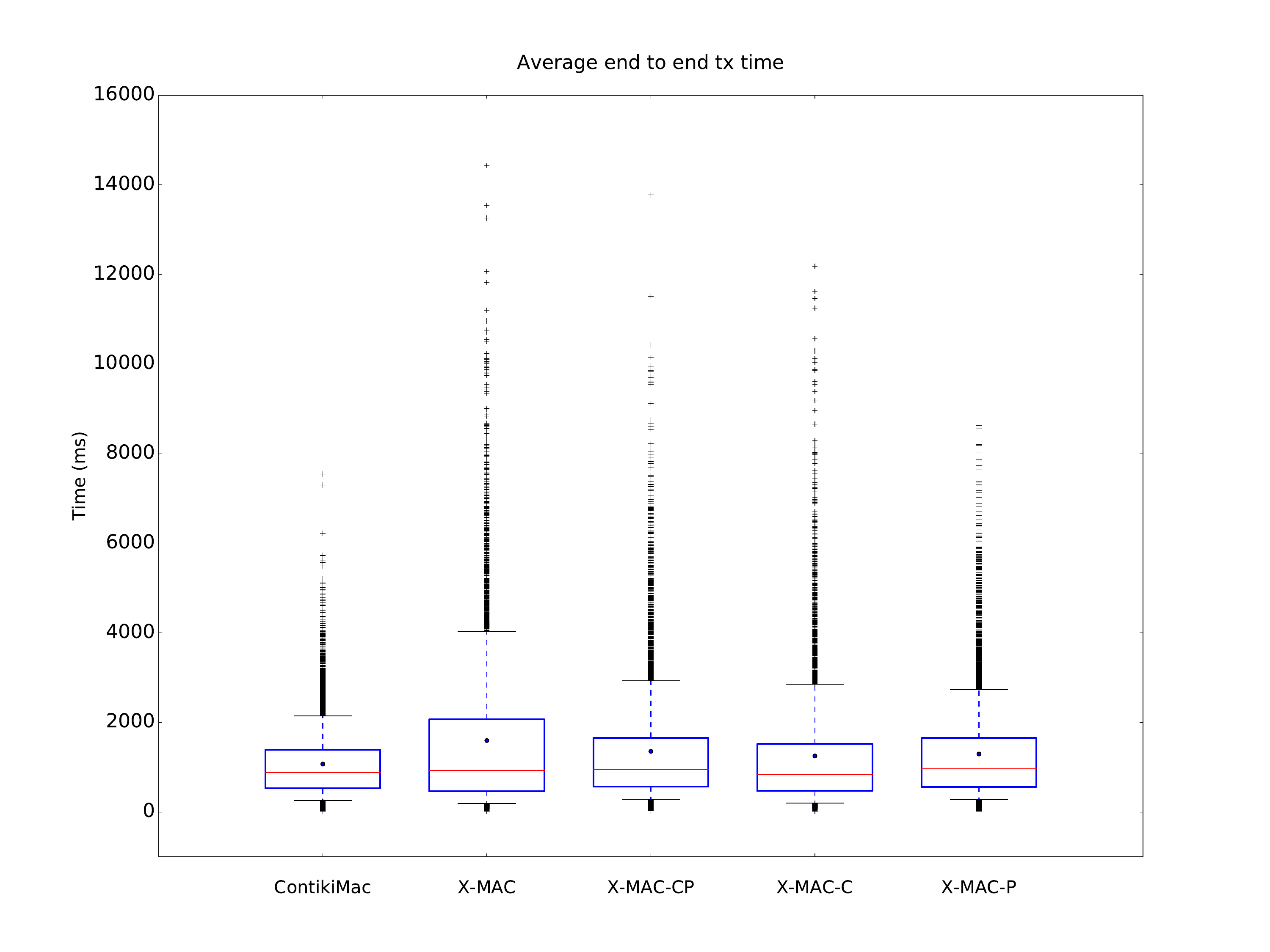}
  \caption{Average end-to-end transmission time.}
 \label{fig:txtime}
 \end{center}
\end{figure}

To further distinguish the behaviour of the different protocols, a
more detailed view of the end-to-end latency is
plotted in Fig.~\ref{fig:txtime}. For each protocol, the box extends from the
25\textsuperscript{th} to the 75\textsuperscript{th} percentiles while
the middle line and the dot represents respectively the median and
mean. The end of the whiskers represent the 10\textsuperscript{th} and
90\textsuperscript{th} percentiles.

Although the median latencies achieved by the different protocols are
similar, some transmissions can be significantly longer. This is
expected as the paths from the sending nodes to the sink have
different hop counts. However, some protocols clearly behave less
efficiently. This is the case for \implem{X-MAC} whose maximum latency (14431 ms) is almost twice as high as the maximum achieved by \implem{ContikiMAC} (7544 ms). Even \implem{X-MAC-C} whose
average latency is the closest to that of \implem{ContikiMAC} has higher maximum value (12179 ms). 

By reducing the number of retransmissions
\implem{X-MAC-CP} can enhance the latency compared to \implem{X-MAC}
but \implem{ContikiMAC} still performs better thanks to the
combination of a lower number of retransmissions with a faster
transmission procedure as explained in
Section~\ref{sec:results-regular-topo}. For 90\%
of the transmissions \implem{ContikiMAC} achieve a latency below 2.1 seconds while \implem{X-MAC} and \implem{X-MAC-CP} respectively achieve latencies below 4 seconds and 2.9s.

%--------------------------------
\subsection{Energy consumption}
\label{sec:results-random-topo-energy}

The energy consumption depends on the time spent by the MCU and radio in their different states. We assume
that negligible energy is consumed when the node sleeps while most energy is
consumed during listening (\textit{LISTEN}), transmitting (\textit{TX})
or receiving (\textit{RX}) states. For this reason, we only focus on the latter states. 
%The \textit{LISTEN} state corresponds to a node
% with its radio transceiver turned on but idle, waiting for an incoming
% transmission; the \textit{TX} state to a node that is actually
% transmitting a frame (strobe or data) while the \textit{RX} state
% corresponds to a node actually busy receiving packets.

Table~\ref{tab:energy} details the average fraction of time
spent by nodes in each of these 3 states. We observe that the fraction
of time spent in \textit{LISTEN} state by \implem{X-MAC} (7.21 \%) is
significantly higher than with \implem{ContikiMAC} (1.38\%). This was 
expected given that the fast sleep optimization of \implem{ContikiMAC}
shortens the wake-up period and allows nodes to reduce their active
period to the minimum. Regardless of which variant of X-MAC is used, the difference
with the \textit{LISTEN} time of \implem{ContikiMAC} remains
considerable even if we can observe a slight enhancement when the
phaselock and collision avoidance mechanisms are introduced.

\begin{table}[h]
\footnotesize
\begin{center}
    \begin{tabular}{|l|l|l|l|l|}
      \hline
       & LISTEN (\%)\space\space & TX (\%)\space\space & RX (\%)\space\space & Total (\%)\\
      \hline
      \textbf{ContikiMAC\space} & 1.38 ($\pm$ 0.36) & 0.33 ($\pm$ 0.14) & 0.07 ($\pm$ 0.08) & 1.78 \\
      \hline
      \textbf{X-MAC} & 7.21 ($\pm$ 1.72) & 0.41 ($\pm$ 0.39) & 0.21 ($\pm$ 0.13) & 7.83 \\
      \hline
      \textbf{X-MAC-CP} & 6.12 ($\pm$ 0.69) & 0.20 ($\pm$ 0.16) & 0.18 ($\pm$ 0.11) & 6.5 \\
      \hline
      \textbf{X-MAC-C} & 6.90 ($\pm$ 1.52) & 0.37 ($\pm$ 0.36) & 0.19 ($\pm$ 0.12) & 7.46  \\
      \hline
      \textbf{X-MAC-P} & 6.24 ($\pm$ 0.72) & 0.22 ($\pm$ 0.16) & 0.18 ($\pm$ 0.12) & 6.64  \\
      \hline
    \end{tabular}
    \end{center}\caption{Average duty cycle statistics}\label{tab:energy}
\end{table}

Surprisingly, \implem{X-MAC} spends a similar
fraction of time transmitting (\textit{TX} state) compared to
\implem{ContikiMAC} despite requiring a higher number of
retransmissions. This can be explained by the difference in their transmission procedures. For \implem{X-MAC}, the time wasted by a failed
transmission attempt might have only been spent in sending short
strobe frames, while for \implem{ContikiMAC} longer data frames have
been sent. The average cost of a retransmission is more important for
\implem{ContikiMAC} than for \implem{X-MAC}.

The introduction of a phaselock mechanism and a CCA-based collision
avoidance mechanism in \implem{X-MAC} allows to reduce the energy consumption. In particular, the\implem{X-MAC-CP} variant spends half as much
time transmitting than \implem{X-MAC}. However, the most advanced
X-MAC variant cannot compete with \implem{ContikiMAC} for what
concerns the total fraction of time spent in an active state.

Finally, looking at the RX column, the fact that \implem{X-MAC}, by the use of a prescribed active period, 
 is more sensitive to overhearing increases the time spent in receiving mode compared to
ContikiMAC where a node can reduce its wake-up period.

%--------------------------------
\subsection{Lesson learned}
\label{sec:results-random-topo-lessons}
  
Our experimentations in realistic scenarios confirm most of the
conclusions drawn in Section~\ref{sec:results-regular-topo}. For the
latency, ETX and energy consumption ContikiMAC remains more
efficient than any X-MAC variant.

For what concerns the packet-delivery ratio, \implem{ContikiMAC}
provides the most reliable transmission. This was not the case in the
experiments on star topologies. We attribute this difference to
the higher likelihood of transmissions involving the same destination
(central node) in star topologies. As explained in
Section~\ref{sec:results-regular-topo-lessons}, this causes synchronization among \implem{ContikiMAC} transmitters.

In the end, \implem{ContikiMAC} provides a lower latency, a lower
energy consumption (lower duty-cycle) and a higher PDR (less
retransmissions) then any X-MAC variant.

% ===================================================================
\section{Conclusion and future works}
\label{sec:conclusion1}

In this paper we performed a detailed analysis of the performance of
ContikiMAC. In particular, we questioned the efficiency of its
transmission procedure that relies on sending full data frames. To
this end, we compared ContikiMAC with X-MAC an older protocol that
first sends smaller strobe frames before sending a data frame. We then
designed several variants of X-MAC in which we added some mechanisms
of ContikiMAC. This methodology allowed us to quantify which
ContikiMAC mechanisms are responsible for the largest performance
improvements. We considered 4 different performance metrics : the
expected transmission count, the end-to-end latency, the packet-delivery
ratio and the energy consumption.

Our experiments have shown than ContikiMAC consistently performs
better than any X-MAC variant for all of the metrics
considered. For latency, retransmissions and PDR, the addition of
phaselock to reduce channel use and the addition of collision
avoidance to reduce collisions allowed X-MAC (\implem{X-MAC-CP}) to
achieve performances closer to that of ContikiMAC. However those
enhancements were not able to reduce X-MAC energy consumption. We
attribute this result to the fast-sleep optimization of ContikiMAC
that allows a node to reduce idle-listening to a minimum. We also
observed that, as expected, retransmissions are more expensive for
ContikiMAC than for X-MAC. However in our scenarios ContikiMAC was
able to compensate this disadvantage by reducing the number of
retransmissions.

The development of X-MAC variants contributed to make X-MAC more
efficient while keeping the same transmission procedure that relies on
sending strobes before a data frame. In the most advanced
variant which includes phaselock and collision avoidance
(\implem{X-MAC-CP}), the number of retransmissions was reduced
considerably. A slight reduction in energy consumption was also
observed, but not enough to compete with ContikiMAC. We believe this
enhanced X-MAC remains an interesting alternative to ContikiMAC as the
latter has been shown to be quite sensitive to the calibration of the CCA
mechanism. In particular, \cite{ccathreshold} have shown that
ContikiMAC can suffer from ``false wake-up'' when the CCA threshold is
not properly calibrated. Environments with high levels of channel
noise could then cause unnecessary node wake-ups and be detrimental to
ContikiMAC's performance.

Finally, we observed that the phaselock mechanism was quite inaccurate
making errors of up to several milliseconds when estimating the
receiving node's wake-up time. Our implementation of phaselock in the
X-MAC variants was even more inaccurate. One possible future work
would then be to study how to improve phaselock accuracy. One possible
approach would be to force nodes to embed an indication of their exact
wake-up schedule in the strobe or data ACK, allowing the sender to
correct its phaselock estimation.

% ======================================================================================================================================

%\bibliographystyle{alpha}
%\bibliography{xmac-contikimac-rpl-llncs}

\bibliographystyle{alpha}
\bibliography{xmac-contikimac-rpl-llncs}

\end{document}